\documentclass[journal]{new-aiaa}
\usepackage[utf8]{inputenc}
\usepackage{textcomp}
\usepackage{graphicx}
\usepackage{epstopdf}
\usepackage{amsmath}
\usepackage[version=4]{mhchem}
\usepackage{siunitx}
\usepackage{svg}
\usepackage{longtable,tabularx}
\usepackage{caption}
\usepackage{multicol}
\usepackage{subcaption,natbib}
\usepackage{soul,xcolor}

\newcommand{\edit}[1]{\textcolor{black}{#1}}

\title{A network-theoretic approach for characterizing\\Mack-mode instability in high-speed boundary layers}

\author{Sai Prasad Mohanty}
\author{Nikhil Khobragade\footnote{Presently at Indian Institute of Technology Madras.}}
\affil{Indian Institute of Science, Bengaluru 560012, India}

\author{Gaurav Chopra\footnote{Presently at Indian Institute of Technology Delhi.}}
\affil{Indian Institute of Technology Madras, Chennai 600036, India}

\author{Aswathi Krishna\footnote{Presently at ISRO Vikram Sarabhai Space Centre.}}
\affil{Indian Institute of Science, Bengaluru 560012, India}

\author{R. I. Sujith}
\affil{Indian Institute of Technology Madras, Chennai 600036, India}

\author{Subrahmanyam Duvvuri\footnote{Email for correspondence: subrahmanyam@iisc.ac.in.}}
\affil{Indian Institute of Science, Bengaluru 560012, India}

\begin{document}

\maketitle

\begin{abstract}
Here we present a network theory-based approach to investigate the Mack-mode instability signature found in high-speed schlieren data from a Mach 6 laminar boundary layer flow over a $7^\circ$ cone. The data contain instability wave packets in the form of coherent rope-like structures which exhibit intermittency. The intermittency implies that conventional Fourier techniques are not particularly well suited for analysis. Network analysis, which is well known for handling episodic spatio-temporal data in a variety of complex systems, provides an alternate and more suitable framework. Techniques from time-varying spatial proximity networks are applied to the present data. The connected components in the network topology reveal lines of constant phase for coherent wave packets associated with the instability, and localized regions of high schlieren light intensity for intermittent \textit{laminar} or \textit{turbulent} flow states. The orientation angle of the connected network components is found to be a suitable metric for identifying components associated with the Mack-mode instability, and that enables detailed characterization of the wavelength and propagation speeds of the instability wave packets. Beyond the characterization exercise, network analysis can provide a powerful framework for understanding the fundamental nature of intermittency and its role in the laminar-to-turbulent flow transition process.
\end{abstract}

\section*{Nomenclature}

{\renewcommand\arraystretch{1.0}
\noindent\begin{longtable*}{@{}l @{\quad=\quad} l@{}}
$A$ & adjacency matrix of a spatial proximity network \\
$A_{ij}$ & element of the adjacency matrix indicating connection between nodes $i$ and $j$ \\
$C_1$ & intensity threshold criterion \\
$C_2$ & spatial proximity criterion \\
$I_i, I_j$ & intensity values of nodes $i$ and $j$, respectively \\
$k$ & wavenumber \\
$M$ & Mach number \\
$N$ & total number of nodes in a network \\
$P$ & pressure \\
$R$ & particular gas constant \\
$T$ & temperature \\
$t$ & time at which a schlieren image is captured \\
$t_1, t_2, t_3$ & arbitrary instants corresponding to schlieren image capture \\
$\Delta t$ & time interval between two consecutive schlieren images \\
$x$ & coordinate in the freestream flow direction \\
$y$ & coordinate normal to the freestream flow direction \\
$s$ & coordinate along the cone surface \\
$n$ & coordinate normal to the cone surface \\
$(s_i, n_i)$ & spatial coordinates of node $i$ in the schlieren image frame \\
$(s_{\text{mean}}, n_{\text{mean}})$ & centroid of a connected component \\
$(\Delta s)_{\text{max}}$ & maximum estimated displacement of a connected component along $s$ \\
$\text{d}(i, j)$ & Euclidean distance between nodes $i$ and $j$ \\
$i, j$ & indices corresponding to specific nodes in the schlieren image \\
$\gamma$ & ratio of specific heats at constant pressure and volume \\
$\Im$ & threshold intensity used to determine active nodes \\
$\Re$ & spatial proximity radius for establishing connections between nodes \\
$\rho$ & density \\[5mm]

\multicolumn{2}{@{}l}{Subscripts}\\
$\infty$ & free stream conditions \\
$0$ & stagnation conditions \\
$e$ & boundary layer edge conditions
\end{longtable*}}

\twocolumn
\section{Introduction}

\lettrine{F}{low} transition from a laminar to a turbulent state in a hypersonic boundary layer can occur through multiple pathways \cite{fedorov2011transition, zhong2012direct}. Among them, the pathway through the second mode linear instability, also known as the Mack-mode instability, is recognized to be important, and is commonly observed in flows over streamlined two-dimensional (2D) or axisymmetric bodies at small angles of attack \cite{mack1975linear,mack1987review,betchov2012stability,criminale2018theory}. Hence, characterizing the disturbances associated with the Mack-mode instability is important for understanding the overall flow transition process. Unlike Tollmien-Schlichting (T-S) waves, which are classified as the first mode instability in Mack's framework and are prevalent in incompressible and compressible boundary layers below Mach 4, the Mack mode arises from an inviscid instability mechanism of acoustic origin \cite{mack1984boundary,mack1987review,reed1996linear}. The instability waves generated by the Mack mode typically possess a characteristic wavelength approximately twice the local boundary layer thickness, and propagate downstream at velocities of the order of boundary layer edge velocity. In practice, this results in oscillation frequencies of the order of several hundred kilohertz. Interestingly, the instability waves exhibit strong temporal intermittency in the flow transition region.

High-speed schlieren visualization is an important experimental tool for studying instabilities in hypersonic boundary layers. The Mack-mode instability imparts a distinctive signature to the density gradient field (which is captured by the schlieren technique) in the form of coherent rope-like structures \cite{laurence2012time,zhang2013hypersonic,laurence2014schlieren,grossir2014hypersonic,kennedy2018visualization,siddiqui2021mack,scholten2022linear}. Traditionally, analysis of such schlieren images has relied on various standard signal processing methods. A notable advancement is the use of these methods in conjunction with the pulse-burst image acquisition technique\cite{laurence2012time, laurence2014schlieren}, which enables the determination of frequency spectra from wavenumber spectra and propagation speeds of instability waves. This approach for determining the frequency spectra overcomes the limitations of high-spatial-resolution cameras that cannot temporally resolve the disturbance frequencies of the order of several hundred kilohertz. The high spatial resolution is needed to ensure that the image scale factor (pixels per meter) exceeds twice the maximum wave number of interest, thus satisfying the Nyquist criterion.

Although traditional signal processing-based approaches have significantly contributed to the understanding of \edit{the} second-mode instability, the development and use of alternative analysis frameworks can offer additional insights into the complex flow dynamics associated with flow transition in hypersonic boundary layers. With this motivation, the present work introduces a \edit{complex networks} approach to analyze schlieren data from a hypersonic boundary layer. While the immediate aim of this effort is to characterize the Mack-mode instability in the flow, it also serves as a demonstration of the utility of \edit{complex networks} in the larger context of understanding hypersonic flow transition.

\edit{Network science, which is rooted in graph theory, is an interdisciplinary field that studies complex networks. It has proven to be valuable in the analysis of complex systems through a framework wherein the components of a system are represented as ``nodes'' and their interactions are identified as ``edges'' \cite{barabasi2013network,newman2018networks,van2010graph}.} This approach has gained traction across various scientific domains, including fluid mechanics, where early applications have predominantly focused on incompressible flows \cite{nair2015network,taira2016network,meena2021identifying}. A variety of network analysis tools are available for studying fluid dynamics problems \cite{taira2022network, iacobello2021review}. Networks for flow visualization data may be constructed from either a Lagrangian or an Eulerian perspective. In the Lagrangian perspective, infinitesimal fluid particles moving along with the flow are considered as nodes. In practice, nodes are assigned to specific flow features that are of interest. For instance, groups of fluid tracers, organized into levels based on their initial wall-normal positions, are represented as nodes, with links established according to the spatial proximity of tracers \citep{iacobello2019lagrangian}. Alternatively, in vortical-interaction networks, free vortices of finite strength are represented as nodes, with links established based on the induced velocity of one vortex on another, as determined by the Biot-Savart law \citep{nair2015network}.  In the Eulerian perspective, nodes are fixed spatial locations where observations are made either using data from experiments or computations. And links are established based on the strength of \edit{the} statistical association of flow activity between nodes.

In the present work we \edit{adopt} the Eulerian perspective and construct time-varying spatial proximity networks from high-speed schlieren data to investigate flow instability characteristics. Pixels in the digital schlieren field of view are taken as nodes, and links are established between nodes in a neighborhood when the nodes simultaneously exhibit high schlieren light intensity. We note that this approach has been used previously to study the spatio-temporal evolution of the laminar separation bubble that forms in a low-speed cylinder flow \cite{chopra2024evolution}, and a similar approach was also used to study the evolution of clusters of acoustic power sources in practical turbulent combustors \cite{krishnan2019emergence}. Application of time-varying spatial proximity networks to the present high-speed schlieren data allows us to estimate the wavelength and velocity of the coherent wave packets arising from Mack-mode instability in the schlieren data. We also make a comparison between the results from the network analysis and the results obtained from application of traditional image processing techniques to the same dataset, and find a good agreement between them. \edit{The utility of network analysis goes beyond the spatio-temporal characterization presented here, and it can serve as a powerful framework for understanding the fundamental nature of intermittency and its role in the overall flow transition process in high-speed boundary layers}.

The rest of this paper is organized in the following manner. Section~\ref{sec:ExperimentalSetup} details the experimental setup, including the test article, schlieren imaging configuration, and pulse-burst imaging technique. Section~\ref{sec:Results} begins by providing an overview of the schlieren data. That is followed by results from traditional image processing methods--wavenumber spectra, propagation velocities, and temporally localized frequency spectra--presented in section~\ref{sec:traditional}. Subsequently, section~\ref{sec:network} details the construction of time-varying spatial proximity networks from schlieren data, followed by a discussion on the connected components of these networks and insights obtained from the analysis. Section~\ref{sec:network} also presents results for identification of Mack mode wave packets, their wavelengths, and convection speeds, thereby demonstrating the utility of the network-based approach as an alternative to traditional analysis methods. Conclusions from this work are presented in section~\ref{sec:conclusions}.

\section{Experimental Setup} \label{sec:ExperimentalSetup}

\subsection{Roddam Narasimha Hypersonic Wind Tunnel}\label{sec:Facility}
Boundary layer flow experiments were performed in the Roddam Narasimha Hypersonic Wind Tunnel (RNHWT) at the Indian Institute of Science. The RNHWT is a 0.5-meter diameter enclosed free-jet facility that operates in pressure-vacuum blowdown mode. It utilizes dry air as the working fluid and can produce free-stream Mach numbers ranging from 6 to 10. All the experimental data presented in this paper was obtained at a free-stream Mach number ($M_\infty$) of 6, with flow stagnation temperature ($T_0$) of $408.7$ K and stagnation pressure ($P_0$) of $12.1$ bar. The resulting unit Reynolds number is approximately $12.7 \times 10^{6} \mathrm{m}^{-1}$. Under these flow conditions the root-mean-square (RMS) value for the free-stream velocity disturbance level in the RNHWT is approximately 0.7\% \cite{thasu2024measurement, premika2024self}.

\subsection{Test Article}
\begin{figure}[hbt!]
    \centering
    \includegraphics[width=3.0in]{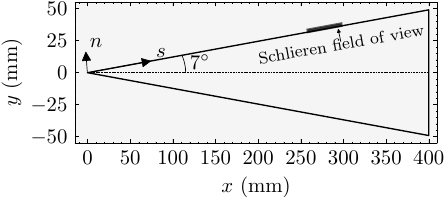}
    \caption{A schematic showing the center plane of the right circular cone model.}
    \label{fig:cone_schematic}
\end{figure}

The model used in this study is a 7\textdegree{} half-angle right circular cone, with a nominally sharp nose tip, at zero angle of attack. A schematic of the model with the coordinate system is shown in Fig~\ref{fig:cone_schematic}.

\subsection{Schlieren}
A Z-type schlieren setup was employed in this study to visualize the flowfield. The rectangular field of view spans a length of 40 mm along the cone surface, starting from $s$ = 260 mm, and has a height of 2 mm. The choice of schlieren location was guided by precursor experiments which showed the boundary layer flow to be transitional in this region. A Cavilux Smart pulsed diode laser with a wavelength of 640 nm and pulse width of 10 ns was used as the light source. The short pulse width enables a high degree of temporal localization in imaging, effectively capturing the fast-moving Mack-mode instability waves without significant motion blur. The knife-edge cut was aligned along the cone surface, and therefore the schlieren light intensity represents density gradients in the wall-normal direction. A Phantom V1612 high-speed camera was used to capture greyscale images. The camera was equipped with a 400 mm f/2.8 Nikon lens, which provides suitable magnification to view the region of interest. The overall schlieren setup is similar to the one used in other recent studies carried out in RNHWT \cite{sasidharan2021large,thasu2022strouhal,kumar2024model}.

\begin{figure*}[hbt!]
    \centering
    \includegraphics[width=0.8\textwidth]{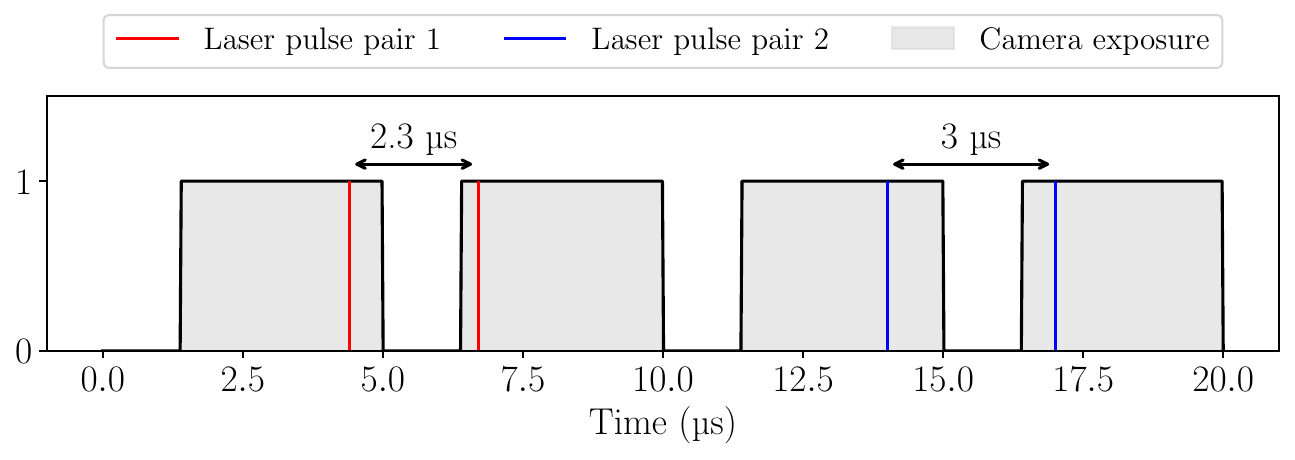}
    \caption{A timing diagram of the camera shutter and light pulses in the experimentally implemented pulse-burst imaging method. Signal level high ($= 1$) indicates an open shutter, and red and blue lines indicate light pulses of 10 ns width. The camera is operated at 200,000 frames per second with an exposure time of \textbf{\SI{3.6}{\micro\second}} for all frames.}
    \label{fig:Laser_timings}
\end{figure*}

Following \citet{laurence2012time,laurence2014schlieren,laurence2016experimental}, the pulse-burst image acquisition technique was implemented for for acquiring the schlieren images. This technique involves illuminating the flowfield at non-uniform time intervals to mitigate aliasing effects in image cross-correlation and to accurately determine the propagation velocity of coherent flow features. The camera image resolution was set to 1152 pixels $\times$ 64 pixels at a fixed frame rate of 200,000 frames per second and an exposure time of $ \SI{3.6}{\micro\second}$ for every frame. To capture images at non-uniform time intervals, the light source is pulsed in a temporal sequence shown in the form of a timing diagram shown in Fig.~\ref{fig:Laser_timings}. Two pairs of light pulses, labeled pair 1 and pair 2, are fired in an alternating sequence. The inter-pulse separation is $ \SI{2.3}{\micro\second}$ and $ \SI{3}{\micro\second}$ for pair 1 and pair 2, respectively. In each pair, the first pulse is timed to occur near the end of a camera gate period, which places the second pulse shortly after the start of the next gate period (the gate period here refers to the camera exposure time). This four-pulse pattern spanning $ \SI{20}{\micro\second}$ is repeated continuously throughout the duration of the experiment. Schlieren data is recorded in packets, where each packet contains a continuous sequence of 2200 images. Multiple such schlieren data packets were acquired, and it was found that separate analysis of each of the data packets yielded identical statistical results.

It is important to note that while network analysis can be applied to data obtained using the pulse-burst imaging technique, it can also be applied to data if it were obtained using a simpler, periodic single-pulse illumination scheme. The former was chosen in the present experiment since it allows for the data to be also processed using conventional Fourier techniques, and thereby a comparison to be made between the two analysis methods.

\section{Results} \label{sec:Results}

\begin{figure*}[hbt!]
    \centering
    \begin{subfigure}[b]{0.82\textwidth}
        \centering
        \includegraphics[width=\textwidth]{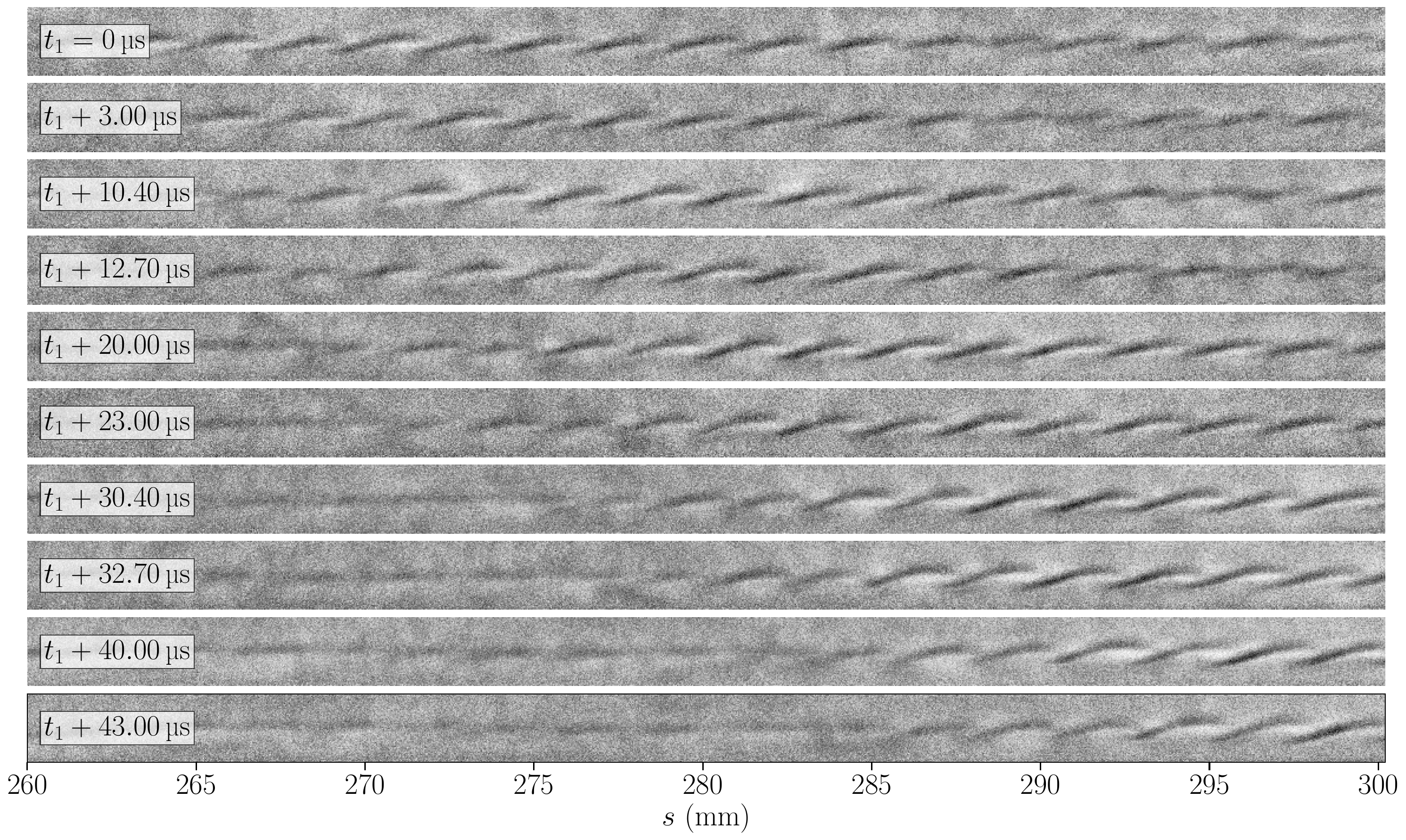}
        \caption{ }
        \label{fig:Mack_mode}
    \end{subfigure}
    \hfill
    \begin{subfigure}[b]{0.82\textwidth}
        \centering
        \includegraphics[width=\textwidth]{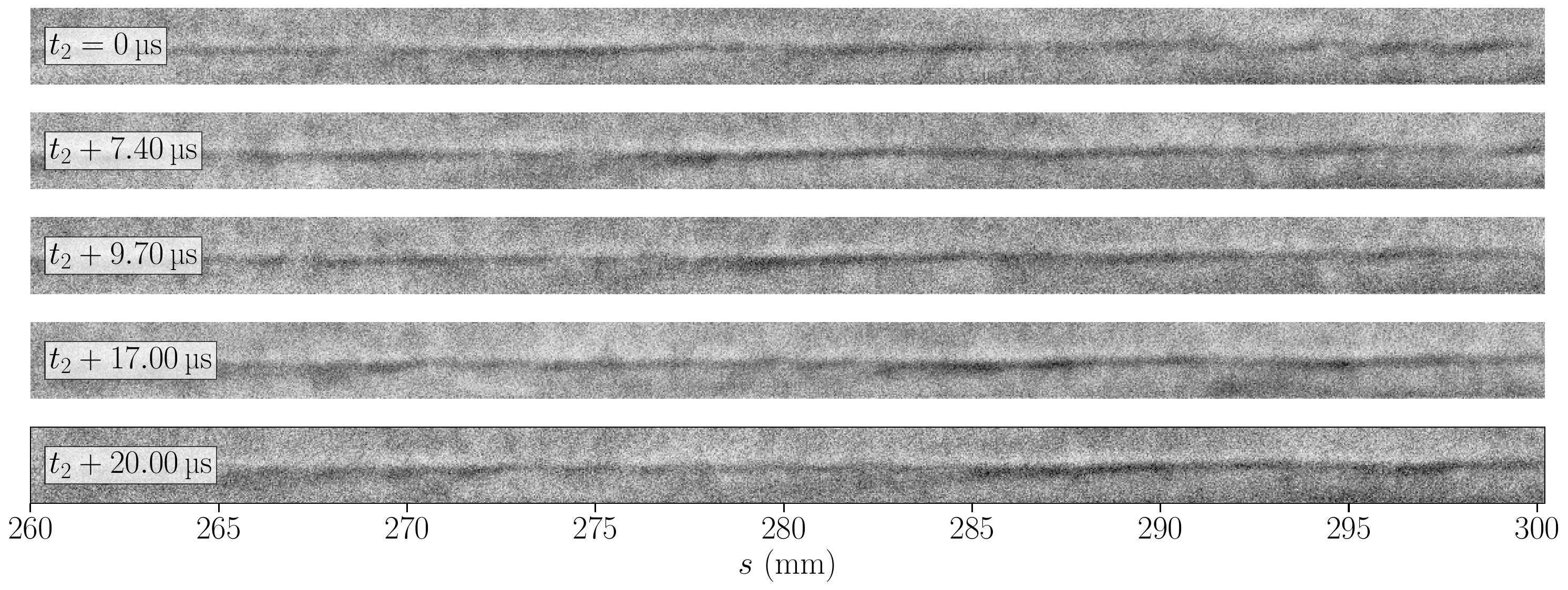}
        \caption{ }
        \label{fig:Laminar_state}
    \end{subfigure}
    \hfill
    \begin{subfigure}[b]{0.82\textwidth}
        \centering
        \includegraphics[width=\textwidth]{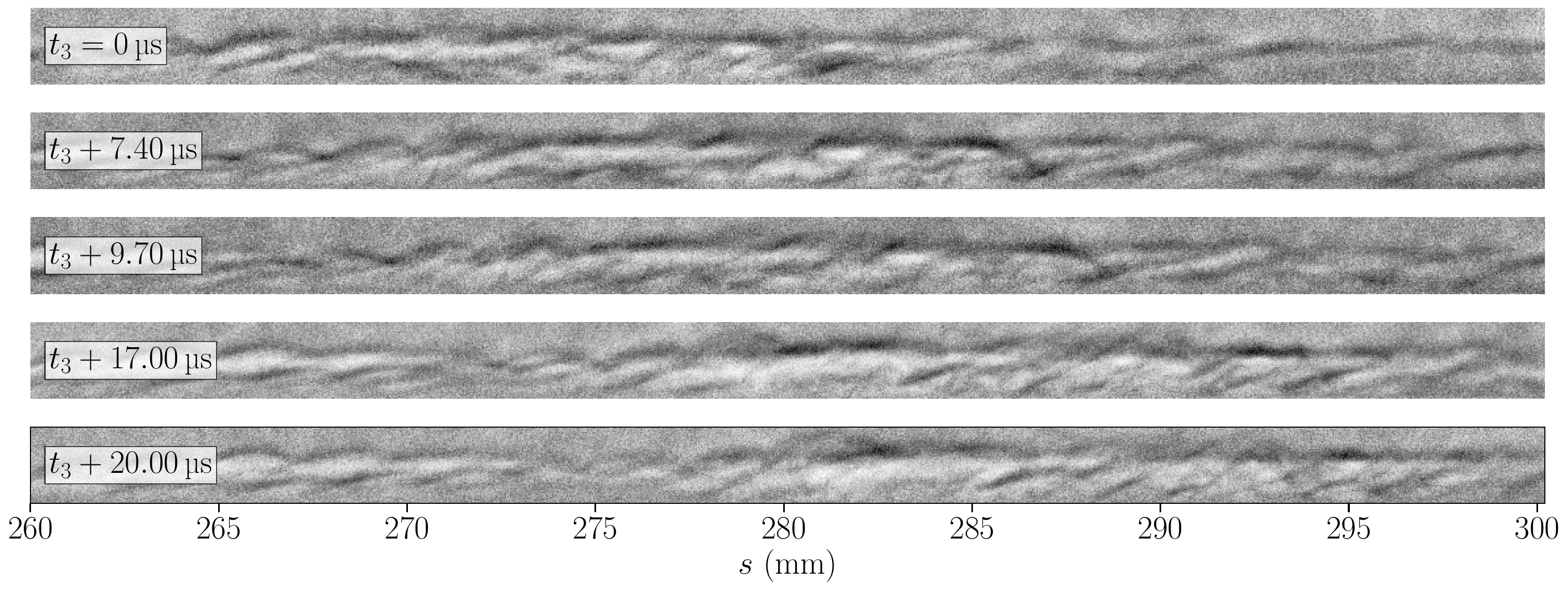}
        \caption{ }
        \label{fig:Onset_of_turbulence}
    \end{subfigure}
    \caption{Schlieren images showing three different flow states: (a) coherent rope-like structures generated by the Mack-mode instability, (b) laminar, and (c) breakdown. All the schlieren image frames in this figure span a height of 2 mm.}
    \label{fig:schlieren_sequence}
\end{figure*}

Mean image subtraction is performed as the first step in the analysis of the schlieren data. All the schlieren images in a data packet are averaged (in time) to obtain a mean schlieren image, and this mean image is then subtracted from each of the individual schlieren images in the data packet. Mean subtraction was done to mitigate the effects of varying light intensity across pulses from the light source \cite{laurence2012time,laurence2014schlieren}. All subsequent analysis is performed on mean-subtracted images, with no other image enhancements. Fig.~\ref{fig:schlieren_sequence} shows three sets of representative mean-subtracted schlieren images from the experiment. The first set in Fig.~\ref{fig:schlieren_sequence}a shows a sequence of ten images, starting at an arbitrary time $t_1$, in which the coherent rope-like disturbances generated by the Mack-mode instability are clearly visible. Downstream motion of the disturbance wave packet can also be discerned from the image sequence. 
These rope-like structures observed in the density gradient field are qualitatively consistent with findings reported earlier in hypersonic boundary layer flow literature \cite{laurence2012time,laurence2014schlieren,laurence2016experimental,zhang2013hypersonic,grossir2014hypersonic,kennedy2018visualization,siddiqui2021mack,scholten2022linear}. In Fig.~\ref{fig:schlieren_sequence}a it is seen that as the disturbance wave packet moves downstream, the region upstream of the wave packet is free of any large-scale coherent disturbances, and the boundary layer locally appears to attain a ``laminar'' state. The second set of schlieren images in Fig.~\ref{fig:schlieren_sequence}b shows the flow in such a laminar state, beginning at an arbitrary time $t_2$.

The coherent flow disturbances grow in strength as they move downstream, and as their growth saturates, they experience a breakdown event, leading to locally ``turbulent'' flow regions. Fig.~\ref{fig:schlieren_sequence}c shows a set of schlieren images of the flow in such a state of breakdown, beginning at an arbitrary time $t_3$. Overall, the boundary layer in the transition region was observed to be highly intermittent, switching between states of laminar flow, growth of coherent rope-like structures, and breakdown of those coherent structures (in that order).  Following these qualitative observations, we now subject the data to quantitative analysis, first using conventional Fourier analysis methods (section~\ref{sec:traditional}), and then using network analysis methods (section~\ref{sec:network}).

\subsection{Fourier analysis of schlieren images} \label{sec:traditional}

\subsubsection{Wavenumber spectra} \label{sec:wavenumber_spectra}
\begin{figure}[hbt!]
    \centering
    \begin{subfigure}[b]{2.65in}
        \centering
        \includegraphics[width=2.6in]{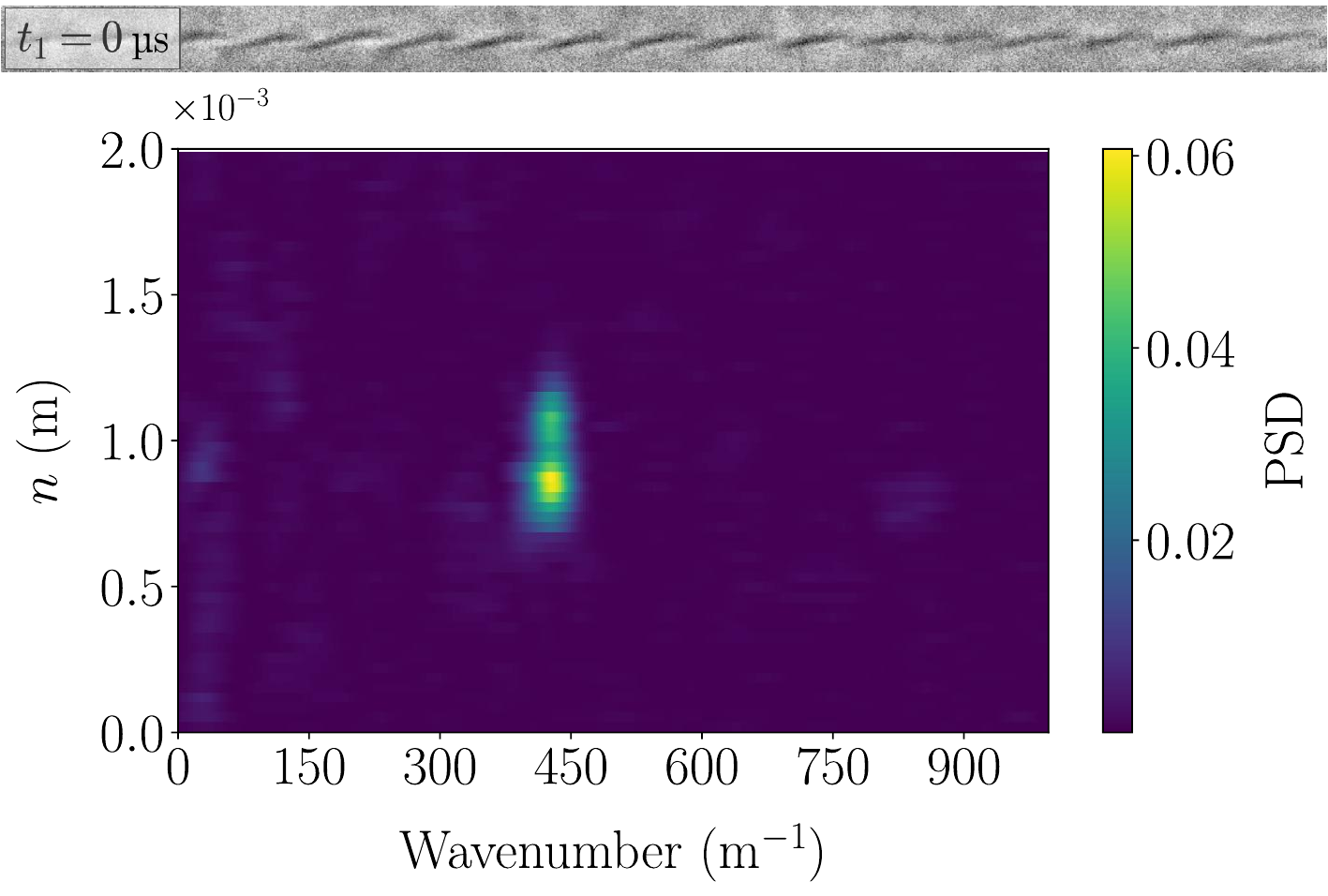}
        \caption{}
        \label{fig:psd1}
    \end{subfigure}
    \hfill
    \begin{subfigure}[b]{2.65in}
        \centering
        \includegraphics[width=2.65in]{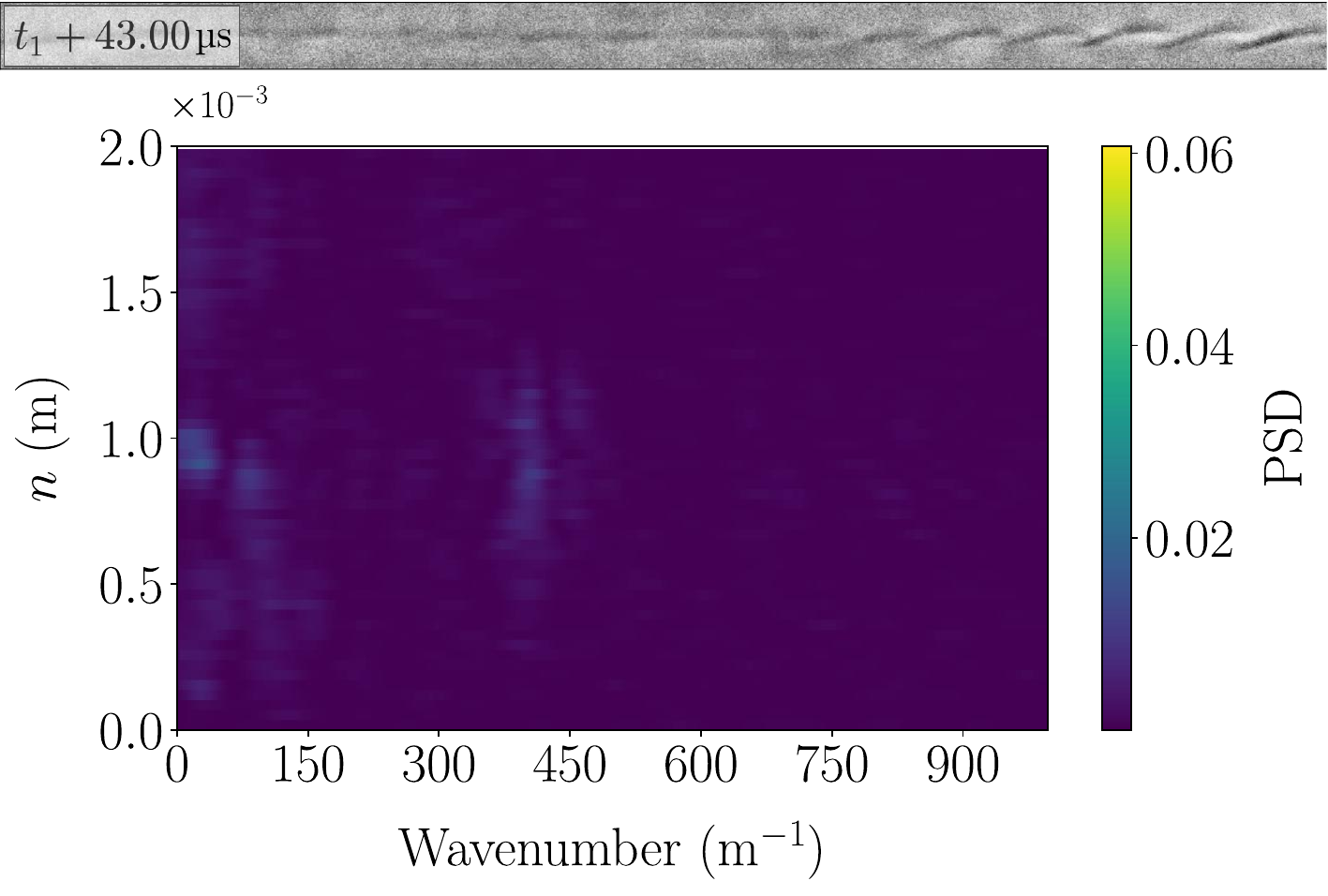}
        \caption{}
        \label{fig:psd3}
    \end{subfigure}
    
    \caption{Wavenumber spectra of the schlieren images displayed in Fig.~\ref{fig:Mack_mode} at two time instants: (a) $\mathbf{t_1 = \SI{0}{\micro\second}}$; (b) $\mathbf{t_1 + \SI{43}{\micro\second}}$.}
    \label{fig:Wave_number_sequence}
\end{figure}

A preliminary understanding of the flow disturbance length scales can be obtained by considering the spatial wavenumber spectra of the schlieren images. Fig.~\ref{fig:Wave_number_sequence} shows the wavenumber power spectral density (PSD) for schlieren intensity fluctuations in the $s$ direction at different $n$ for two specific image frames shown in Fig.~\ref{fig:Mack_mode}. At \(t_1 = \SI{0}{\micro\second} \) (Fig.~\ref{fig:psd1}), a distinct peak in the spectra is observed at wavenumber $k\approx 430$ m$^{-1}$, at a height of $n\approx 1\,\text{mm}$ from the surface where the wave packets associated with the Mack-mode instability are centered. This wavenumber corresponds to a wavelength of approximately 2.3 mm. At \(t_1 + \SI{43}{\micro\second} \) (Fig.~\ref{fig:psd3}), the wave packet has moved downstream and the flow in the majority region of the field of view is laminar. As expected, the clear spectral signature of the Mack-mode instability seen in (Fig.~\ref{fig:psd1}) is now missing, since the wave packet spans (in $s$) only a relatively short extent of the field of view. We note in passing that in this scenario the Fourier basis not the optimal basis to decompose the coherent fluctuations since the wave packet does not span the entire data window.

\subsubsection{Wave propagation speeds} \label{sec:wave_propogation_speeds}
\begin{figure}[hbt!]
    \centering
    \includegraphics[width=2.6in]{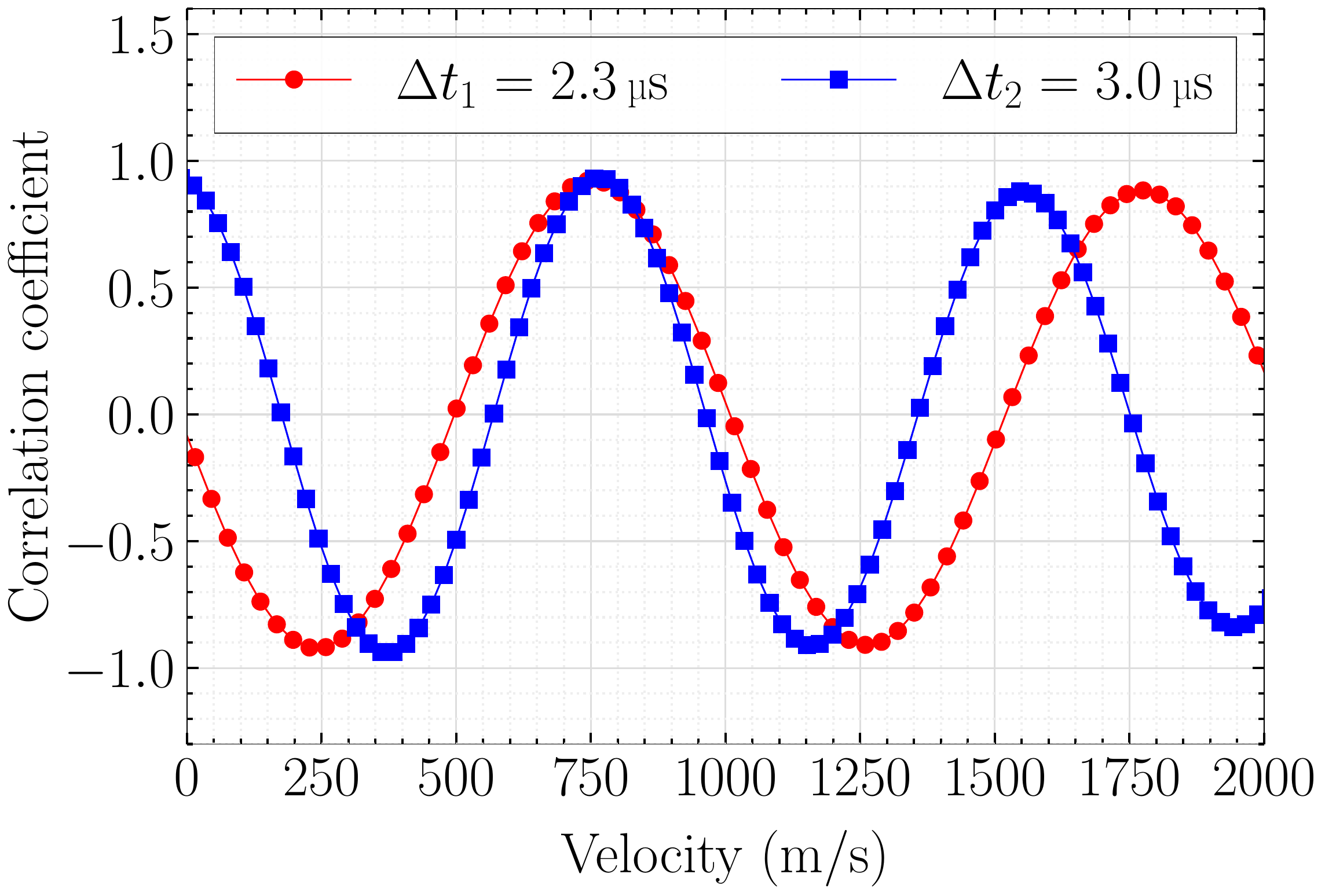}
    \caption{Correlation coefficients at different propagation speeds obtained for two image pairs in the schlieren image sequence shown in Fig.~\ref{fig:Mack_mode}. Pair-1: images at \( t_1 + \SI{10.4}{\micro\second} \) and \( t_1 + \SI{12.7}{\micro\second} \). Pair-2: images at \( t_1 + \SI{20.0}{\micro\second} \) and \( t_1 + \SI{23.0}{\micro\second} \).}
    \label{fig:cross-corr}
\end{figure}

Image cross-correlation techniques are employed to estimate the propagation speed of the wave packets. Two consecutive image pairs were selected for this purpose: pair-1, with a time interval of \SI{2.3}{\micro\second} between the pair of images, and pair-2, with a time interval of \SI{3.0}{\micro\second} between the pair of images. A spatial bandpass filter, centered around the dominant wavenumber of the Mack-mode instability, was applied to images in each pair to accentuate the most prominent features of the wave packet. Subsequently, cross-correlation in the streamwise direction was computed between the images in each pair to determine the correlation values as a function of displacement in the streamwise direction. These displacements are then converted into velocities using the time interval between the two images of each pair. As a representative example, consider pair-1 to consist of the 3rd and 4th images from the schlieren image sequence shown in Fig.~\ref{fig:Mack_mode} (with timestamps \(t_1 + \SI{10.4}{\micro\second}\) and \(t_1 + \SI{12.7}{\micro\second} \)), and pair-2 to consist of the 5th and 6th images from the same figure (with timestamps \( t_1 = \SI{20.0}{\micro\second} \) and \( t_1 = \SI{23.0}{\micro\second} \)). The cross-correlation coefficient curves for these images pair are presented in Fig.\ref{fig:cross-corr}. For both pair-1 and pair-2, multiple peaks are observed in the cross-correlation coefficient curves, where only one of the peaks corresponds to the true wave packet velocity, and the other secondary peaks appear due to the spatially periodic nature of coherent disturbances (an aliasing effect). The two correlation coefficient curves will show a common peak only at the true propagation velocity for the wave packet. From Fig.\ref{fig:cross-corr}, the common peak is readily identified to be at a velocity of 749.5 m/s, which is taken to be the true propagation velocity of the wave packet that is seen in the two image pairs. It is noted that having two image pairs with different time intervals between the images contained by those pairs allows us to separate the true velocity peak from the aliasing peaks in the correlation coefficient curves, and this forms the basis for the pulse burst imaging technique.

\begin{figure}[hbt!]  
    \centering
    \includegraphics[width=2.6in]{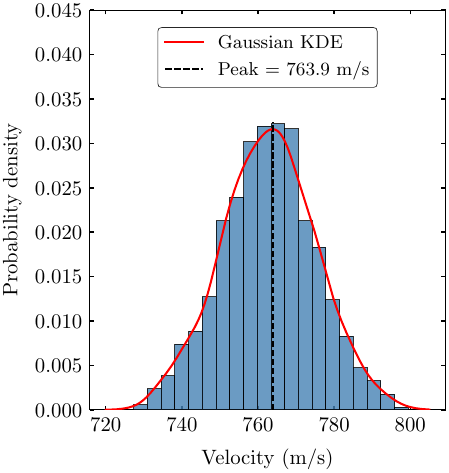}
    \caption{A histogram of the propagation velocities obtained from all image pairs contained in the schlieren data.}
    \label{fig:Histogram_Cross_corr_Velocity}
\end{figure}

This process of identifying the propagation velocity was carried out for all the image sets in the available schlieren data, where each set consists of two successive image pairs (with time intervals of \SI{2.3}{\micro\second} for the first pair and \SI{3}{\micro\second} for the second pair). The propagation velocities obtained from this exercise are shown in the form of a histogram in Fig.~\ref{fig:Histogram_Cross_corr_Velocity}. The probability density function (PDF) exhibits a peak at a velocity of \( 763.9 \, \text{m/s} \), which is about 87\% of the freestream velocity. The PDF gives a mean velocity of \( 762.7 \, \text{m/s} \) and a standard deviation of \( 12.2 \, \text{m/s} \). Given the symmetry of the velocity distribution and the narrow standard deviation, the peak of the PDF closely aligns with the mean, suggesting that both metrics are consistent indicators of the propagation velocity of the Mack-mode instability waves captured in the schlieren data.

\subsubsection{Temporally-local frequency spectra}\label{sec:TLFS}

\begin{figure*}[h!]
    \centering
    \includegraphics[width=\textwidth]{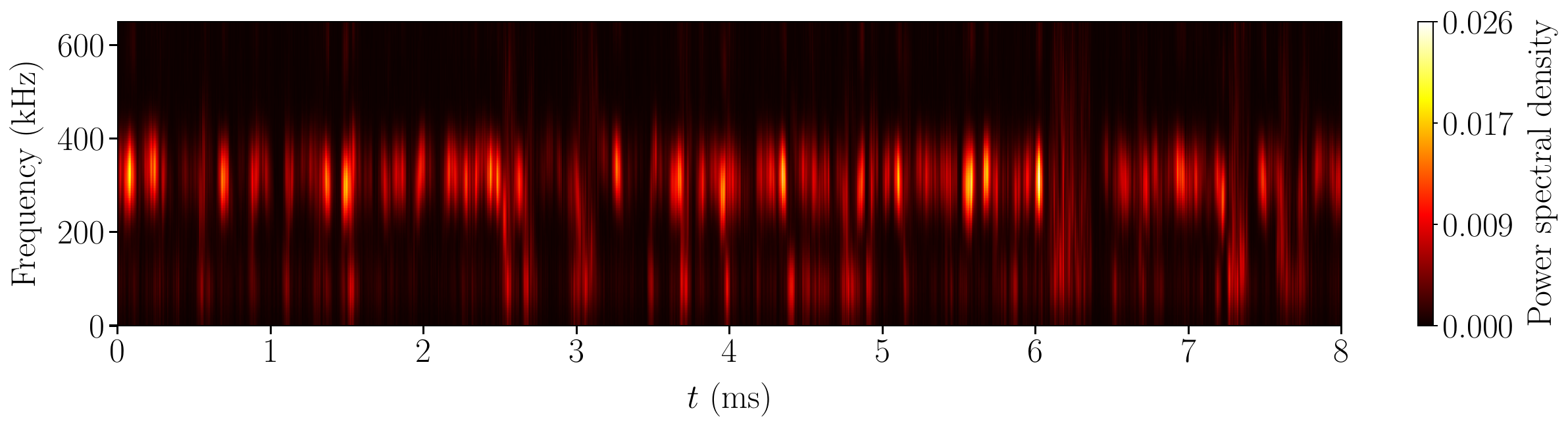}
    \caption{Temporally-local frequency spectra.}
    \label{fig:Temporally-local_frequency_spectra}
\end{figure*}

Temporally-local frequency spectra was obtained by combining the wavenumber spectrum (sec.~\ref{sec:wavenumber_spectra}) with the propagation speeds of wave packets (sec.~\ref{sec:wave_propogation_speeds}). For each image, the row of pixels with the highest signal magnitude was identified, and its wavenumber spectrum was computed. This wavenumber spectrum was then multiplied by the phase speed of the Mack-mode instability waves (ascertained to be \(763.9 \, \text{m/s}\) from Fig.~\ref{fig:Histogram_Cross_corr_Velocity}) to obtain the frequency spectrum for that image. Repeating this process for all images in the schlieren dataset produces a series of time-stamped frequency spectra, wherein each of the spectral curves corresponds to the time instant at which that particular was captured. These spectra are presented as a colormap in Fig.~\ref{fig:Temporally-local_frequency_spectra}. The spectra reveals three flow characteristics:
\begin{itemize}
    \item[(i)] Instances where the spectral peak occurs within the range of 300–350 kHz correspond to the presence of Mack-mode instability.
    \item[(ii)] Instances where the peak shifts to a lower frequency or displays a broadband energy distribution indicate the breakdown of the Mack-mode instability waves.
    \item[(iii)] Instances with no discernible energy across the frequency band or only a faint signal near the lower frequency limit correspond to a disturbance-free laminar flow state.
\end{itemize}
Following this time-series analysis, the remainder of the paper presents results from complex network analysis. And then a comparison is made between the results obtained from time-series analysis and complex network analysis.

\subsection{Network analysis} \label{sec:network}
\begin{figure*}[hbt!]
    \centering
    \includegraphics[width=5.5in]{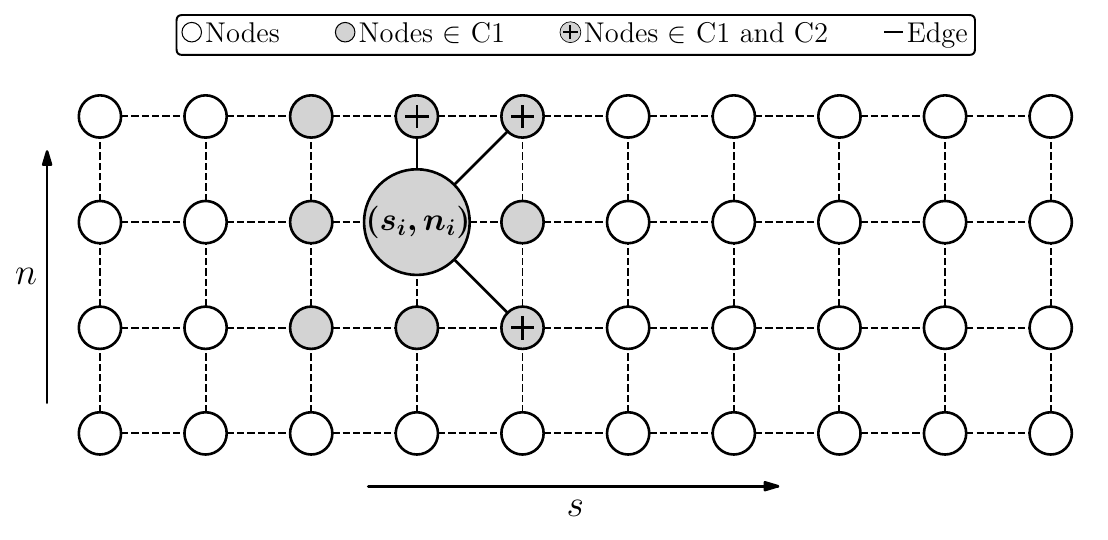}
    \caption{A schematic illustration of the spatial proximity network.}
    \label{fig:network_construction}
\end{figure*}

To construct time-varying spatial proximity networks from a time series of schlieren images, we build a spatial proximity network for each image. In this framework, each pixel in a given schlieren image is treated as a node, and connections (edges) between nodes are established based on two conditions:
\begin{enumerate}
    \item Intensity Threshold (C1): The intensity values of two nodes must both exceed a predefined threshold, $\Im$, for them to be considered for a connection.
    \item Spatial Proximity (C2): The distance between these two nodes must be less than or equal to a specified radius, $\Re$.
\end{enumerate}
The schematic in Fig.~\ref{fig:network_construction} illustrates these conditions in a visual manner, where each pixel in a given schlieren image, represented by circles, is treated as a node. To elaborate on the conditions for connecting nodes (C1 and C2), consider a node indexed \( i \), located at position \( (s_i, n_i) \) with intensity \( I_i \) that exceeds the threshold \( \Im \) (i.e., \( I_i > \Im \)). All pixels within a circular area of radius \( \Re \), centered at \( (s_i, n_i) \), are identified as potential neighbors, satisfying the spatial proximity condition (C2). Among these neighbors, only those whose intensity values also exceed \( \Im \) are connected to the node \( i \) with an edge. These connected nodes are marked with a ``+'' sign in their gray-shaded circles. Repeating this process for all nodes in the image results in the construction of a complete spatial proximity network.

Mathematically, a network is represented by an adjacency matrix \( \mathbf{A} \), which encodes the connections between the nodes in the network. This matrix \( \mathbf{A} \) has dimensions \( N \times N \), where \( N \) is the total number of nodes. In this case, since each pixel in the schlieren image is treated as a node, \( N \) corresponds to the total number of pixels in the image. Each row and column in \( \mathbf{A} \) reflects the connections of a specific node. For any node indexed \( i \), the entries in row \( i \) (or column \( i \)) indicate its connections to other nodes. An entry of \( 0 \) indicates no connection, while an entry of \( 1 \) indicates a connection exists. Therefore, any entry \( A_{ij} \) denotes whether there is a connection between node \( i \) and node \( j \), defined as follows:
\[
A_{ij} = 
\begin{cases} 
1 & \text{if } \text{d}(i, j) \leq \Re \text{ and } I_i \geq \Im \text{ and } I_j \geq \Im, \\ 
0 & \text{otherwise}.
\end{cases}
\]
where \( \text{d}(i, j) \) is the Euclidean distance between nodes \( i \) and \( j \), and \( I_i \) and \( I_j \) are the intensity values of nodes \( i \) and \( j \), respectively.

\begin{figure}[hbt!]
    \centering
    \includegraphics[width=2.6in]{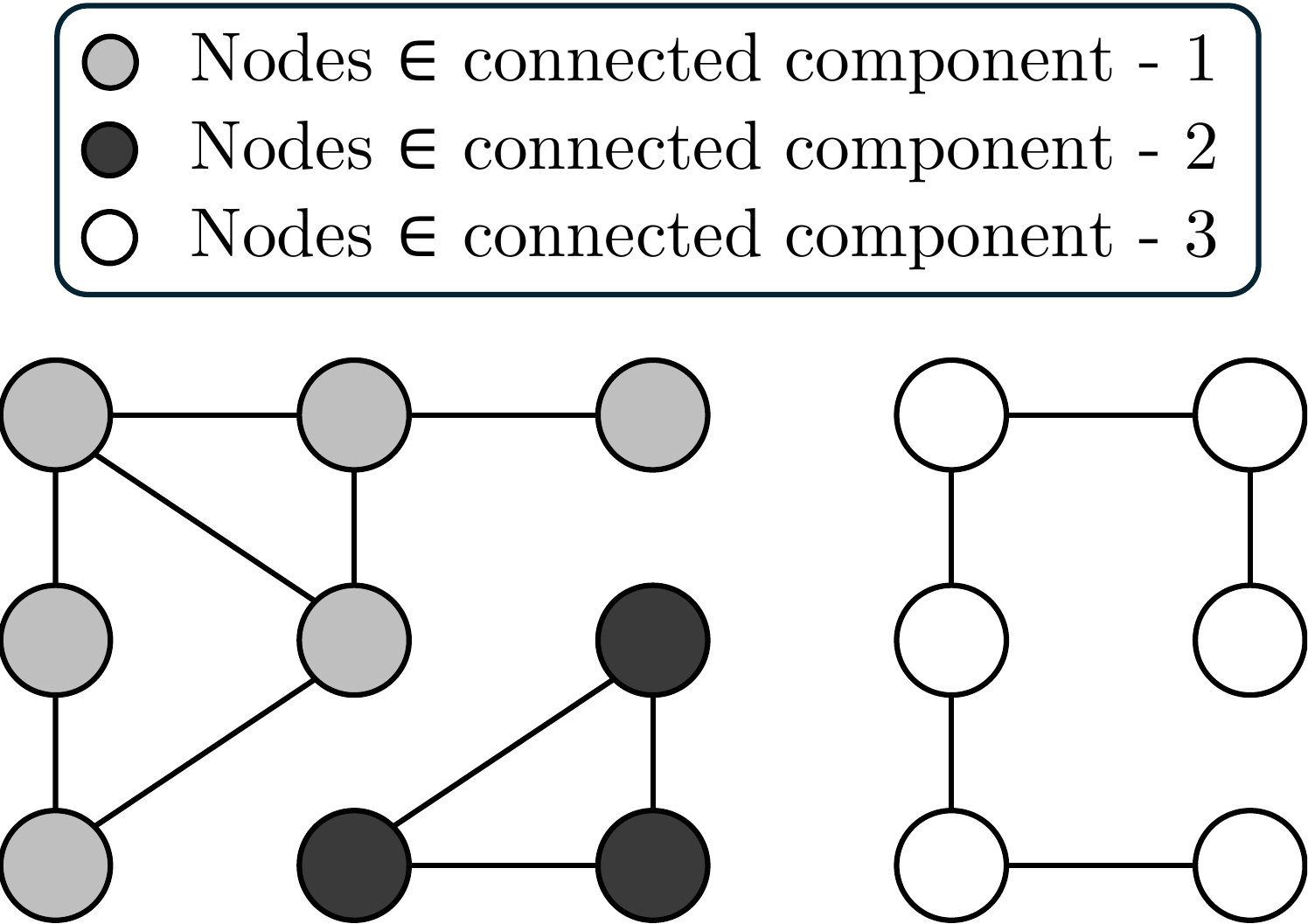}
    \caption{Schematic representation of connected components within a network.}
    \label{fig:connected_components}
\end{figure}

A network can consist of multiple connected components, with each component representing a distinct subnetwork. A connected component is a maximal set of nodes where every pair of nodes is connected, either directly or via a sequence of edges, and no node in one component is connected to a node in another. This results in completely disjoint components, each forming an isolated subnetwork and partitioning the network into discrete clusters or islands \cite{barabasi2013network,newman2018networks}. This concept is illustrated in Fig.~\ref{fig:connected_components}, which provides a schematic representation of connected components in a network.

Identifying the connected components of a network can be highly beneficial, particularly when the network is constructed from an Eulerian perspective using flow visualization data, as connected components highlight critical flow features and facilitate the isolation of these features from other aspects of the flow. Depending on the type of data used for network construction and the methodology employed, connected components can yield various interpretations \cite{chopra2024evolution, krishnan2019emergence}. In the context of the present study, the connected components of the spatial proximity network derived from a schlieren image in the transitional region of the hypersonic boundary layer may indicate either the line of constant phase of Mack-mode instability or structures related to laminar flow and flow breakdown. Further details on these interpretations and their underlying basis will be presented along with the results. 

\begin{figure*}[hbt!]
    \centering
    \begin{subfigure}[t]{0.9\textwidth}
        \centering
        \includegraphics[trim=0cm 31.02cm 0cm 0cm, clip, width=\textwidth]{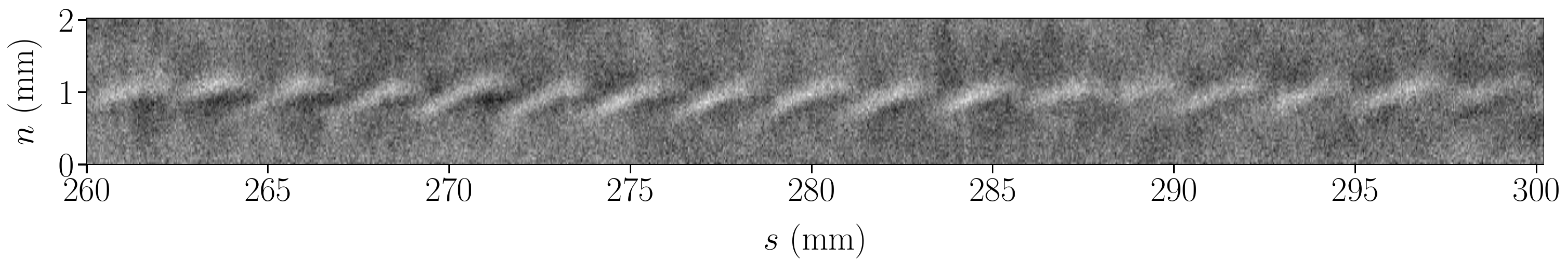}
        \caption{}
        \label{fig:sub1}
    \end{subfigure}

    \begin{subfigure}[t]{0.9\textwidth}
        \centering
        \includegraphics[width=\textwidth]{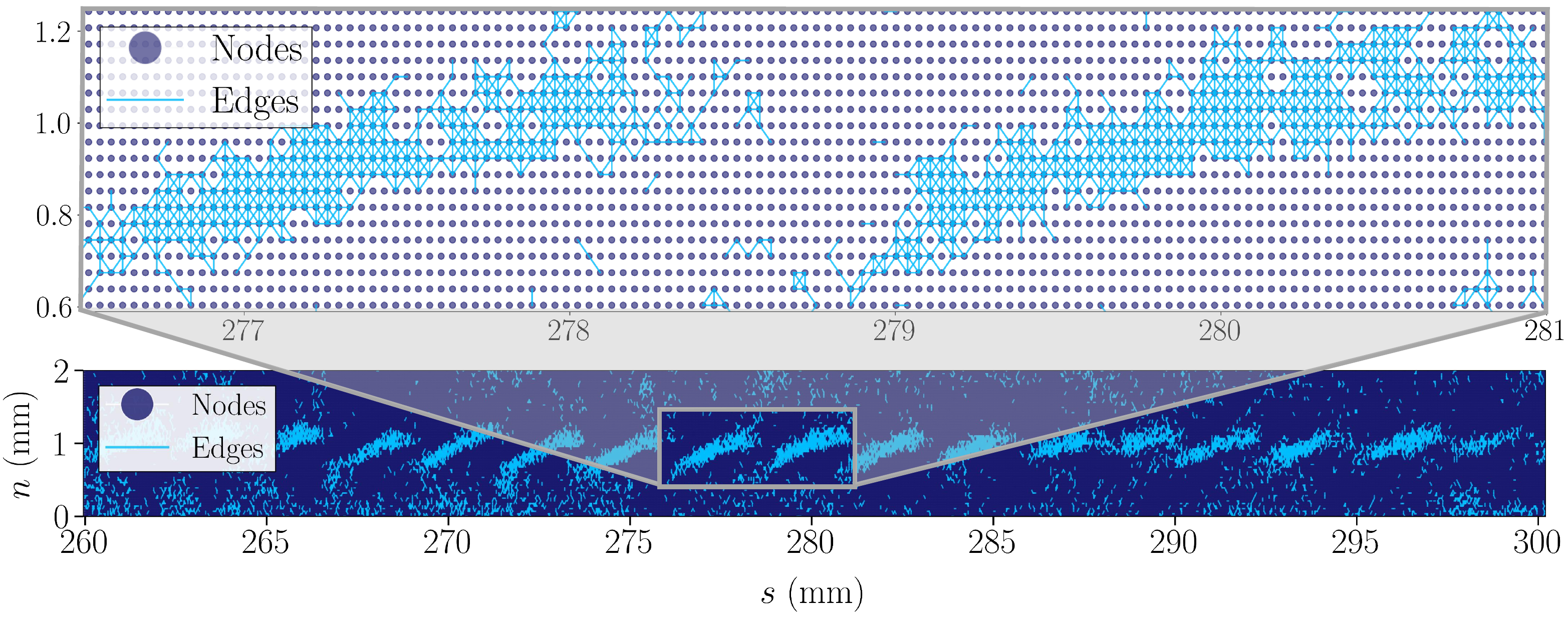}
        \caption{}
        \label{fig:Net_sub}
    \end{subfigure}

    \begin{subfigure}[t]{0.9\textwidth}
        \centering
        \includegraphics[trim=0cm 17.7cm 0cm 9.15cm, clip, width=\textwidth]{Mack_mode_Img_866.pdf}
        \caption{}
        \label{fig:sub2}
    \end{subfigure}

    \begin{subfigure}[t]{0.9\textwidth}
        \centering
        \includegraphics[trim=0cm 0cm 0cm 21.9cm, clip, width=\textwidth]{Mack_mode_Img_866.pdf}
        \caption{}
        \label{fig:sub3}
    \end{subfigure}

    \caption{A representative spatial proximity network constructed for a schlieren image. (a) Schlieren image showing the Mack-mode instability, (b) Spatial proximity network constructed from the image in Fig.~\ref{fig:sub1}, (c) Node degree values derived from the network in Fig.~\ref{fig:Net_sub}, and (d) Top 15 connected components based on the node count.}
    \label{fig:overall}
\end{figure*}

\edit{A spatial proximity network can be constructed for any given schlieren image by following the procedure outlined above. As an representative example, we describe here the network construction process for the first image from the schlieren sequence shown in Fig.~\ref{fig:Mack_mode} (which is reproduced again in Fig.~\ref{fig:sub1} for easy reference). To begin with, all pixels in the image are ranked in descending order based on their light intensity values. The mean intensity of the top 30,000 pixels is then defined as the intensity threshold, \( \Im \), for the image under consideration. To determine network connectivity, a spatial proximity threshold, \( \Re \), is defined based on the Moore neighborhood which comprises of the eight nodes immediately adjacent to a given node in the horizontal, vertical, and diagonal directions.}
The resulting spatial proximity network is presented in Fig.\ref{fig:Net_sub}. Once the network is constructed, the degree of each node (or pixel) can be readily determined. The degree, which is a crucial property of a node, is defined as the number of nodes connected to a given node through one or more edges. Fig.~\ref{fig:sub2} displays a heat map of node degree values derived from the network, where each node's degree is represented by colour while preserving the spatial arrangement of nodes as in the original schlieren image. This visualization effectively reveals local connectivity patterns, highlighting prominent flow features. In this case, coherent disturbances in the flow field stand out distinctly from the rest of the background. By analyzing the connected components of the network, these disturbances can be further isolated from the rest of the flow field. Fig.~\ref{fig:sub3} shows the top 15 connected components, based on the number of nodes, with each component represented in a distinct colour and all nodes belonging to the same component shown in the same colour.

\begin{figure*}[hbt!]
    \centering
    \begin{subfigure}[t]{\textwidth}
        \centering
        \includegraphics[trim=0cm 26.2cm 0cm 0cm, clip, width=\textwidth]{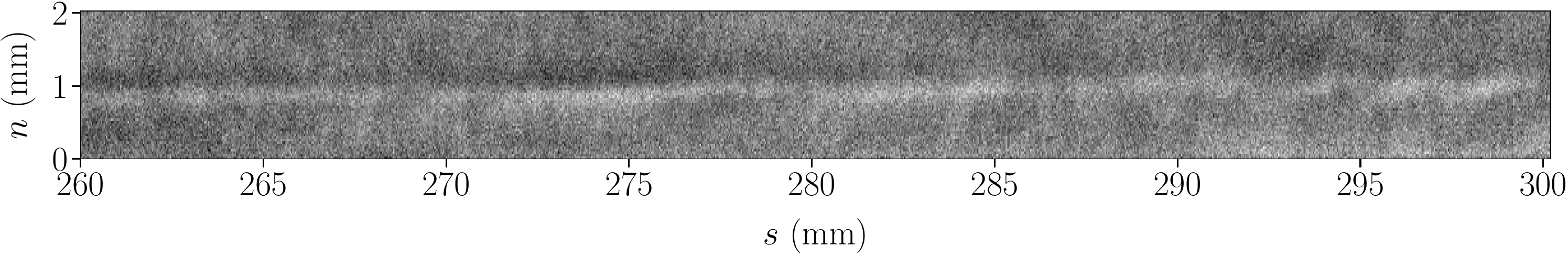}
        \caption{}
        \label{fig:Laminarsub1}
    \end{subfigure}
    \vspace{1em} 
    \begin{subfigure}[t]{\textwidth}
        \centering
        \includegraphics[trim=0cm 0cm 0cm 21.8cm, clip, width=\textwidth]{Laminar_Image.pdf}
        \caption{}
        \label{fig:Laminarsub2}
    \end{subfigure}
    \caption{(a) Schlieren image and (b) corresponding top 15 connected components showing the undisturbed flow field.}
    \label{fig:Laminar_figure}
\end{figure*}

\begin{figure*}[hbt!]
    \centering
    \begin{subfigure}[t]{\textwidth}
        \centering
        \includegraphics[trim=0cm 26.2cm 0cm 0cm, clip, width=\textwidth]{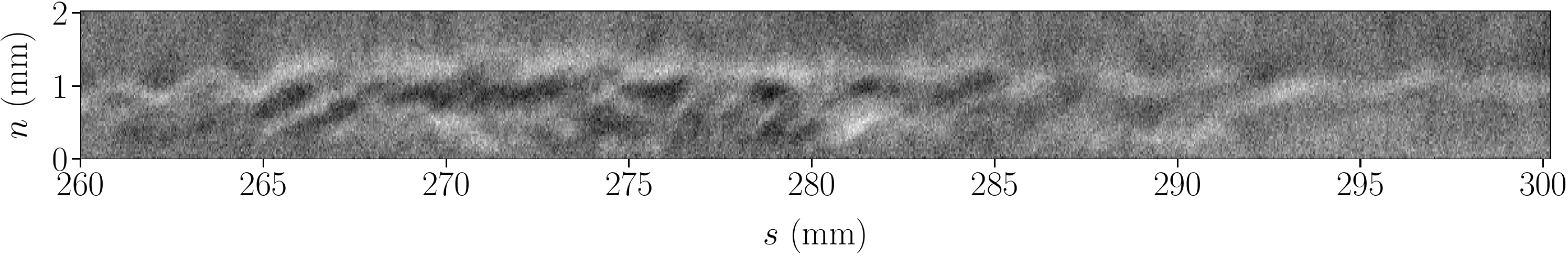}
        \caption{}
        \label{fig:Turbsub3}
    \end{subfigure}
    \vspace{1em} 
    \begin{subfigure}[t]{\textwidth}
        \centering
        \includegraphics[trim=0cm 0cm 0cm 21.8cm, clip, width=\textwidth]{Turbulent_Img.pdf}
        \caption{}
        \label{fig:Turbsub4}
    \end{subfigure}
    \caption{(a) Schlieren image and (b) corresponding top 15 connected components showing flow breakdown.}
    \label{fig:Turbulent_figure}
\end{figure*}

A comparison of Figs.~\ref{fig:sub2} and \ref{fig:sub3} with the schlieren image in Fig.~\ref{fig:sub1} demonstrates how effectively the disturbances are highlighted in Figs.~\ref{fig:sub2} and isolated from the rest of the flow field in Fig.~\ref{fig:sub3}. This was achieved by carefully tuning the values of the parameters \( \Im \) and \( \Re \) in constructing the network. \edit{A small increase from the Moore neighborhood definition for \( \Re \), while keeping \( \Im \) fixed, does not significantly alter the network structure. However, increasing the spatial radius substantially results in a densely connected network with very few connected components, each containing a large number of nodes (on the order of the total network size). In such a scenario almost the entire network forms a single large connected component, making it impossible to distinguish localized disturbances from the rest of the flow field. Conversely, reducing \( \Re \) to a slightly smaller spatial radius corresponding to the von Neumann neighborhood does not lead to a significant change. However, as expected, further reduction in \( \Re \) results in a network with virtually no connections. Therefore, selecting \( \Re \) to match the Moore neighborhood was found to be best suited for the present analysis.} With this value of \( \Re \), it was found that a slight change in \( \Im \), for instance, adjusting it to the mean intensity of the top 25,000 or 35,000 pixels, does not result in a major change. As a result, the node degree map and the corresponding components remain almost identical to those shown in Figs.~\ref{fig:sub2} and \ref{fig:sub3}. However, a significant increase in \( \Im \) results in a disconnected sparse network, while a significant decrease in \( \Im \) leads to a densely connected network with large connected components, where disturbances appear merged rather than distinctly separated, as seen in Fig.~\ref{fig:sub3}. With this understanding, the spatial proximity networks constructed for every schlieren image in our analysis use an \( \Im \) value corresponding to the mean intensity of the top 30,000 pixels in that image and an \( \Re \) value \edit{consistent with the Moore neighborhood definition.}

Using these parameter values, a network was constructed for the first image in the sequence shown in Fig.~\ref{fig:Laminar_state} (reproduced in Fig.~\ref{fig:Laminarsub1}), an instant when the flow field is free of disturbances, and for the first image in the sequence shown in Fig.~\ref{fig:Onset_of_turbulence} (reproduced in Fig.~\ref{fig:Turbsub3}), an instant when flow breakdown is observed. The corresponding connected components derived from each network are shown in Fig.~\ref{fig:Laminarsub2} and Fig.~\ref{fig:Turbsub4}, respectively. It should be emphasized that the interpretation of the connected components in Figs.~\ref{fig:sub3}, \ref{fig:Laminarsub2}, and \ref{fig:Turbsub4} vary. In Fig.~\ref{fig:sub3}, the connected components highlight disturbances in the flow field, with each component aligning along a line of constant phase associated with the Mack-mode instability. In contrast, in the other two figures, Figs.~\ref{fig:Laminarsub2} and \ref{fig:Turbsub4}, the interpretation is less straightforward. Here, we can generally state that the connected components capture localized regions of high pixel intensity in the schlieren image. These regions may correspond to structures associated with a laminar flow field in the former case or a turbulent flow field in the latter.

Given our interest in characterizing the disturbances associated with the second-mode instability, it is important to establish a criterion or metric to identify components specifically related to this instability, i.e., those that align with the line of constant phase. The orientation angle of a connected component can serve as a statistically suitable metric to evaluate whether a component corresponds to the second-mode instability. Earlier studies have shown that the local orientation angle is typically in the range of \(11^\circ\) to \(15^\circ\) around the high-activity zone, i.e., \(y/\delta = 0.6 - 0.8\), where \(\delta\) represents the boundary layer thickness \cite{parziale2015observations,laurence2016experimental}. Thus, the average angle of the second-mode structures is expected to fall within the same range.

\begin{figure}[hbt!]
    \centering
    \includegraphics[width=2.6in]{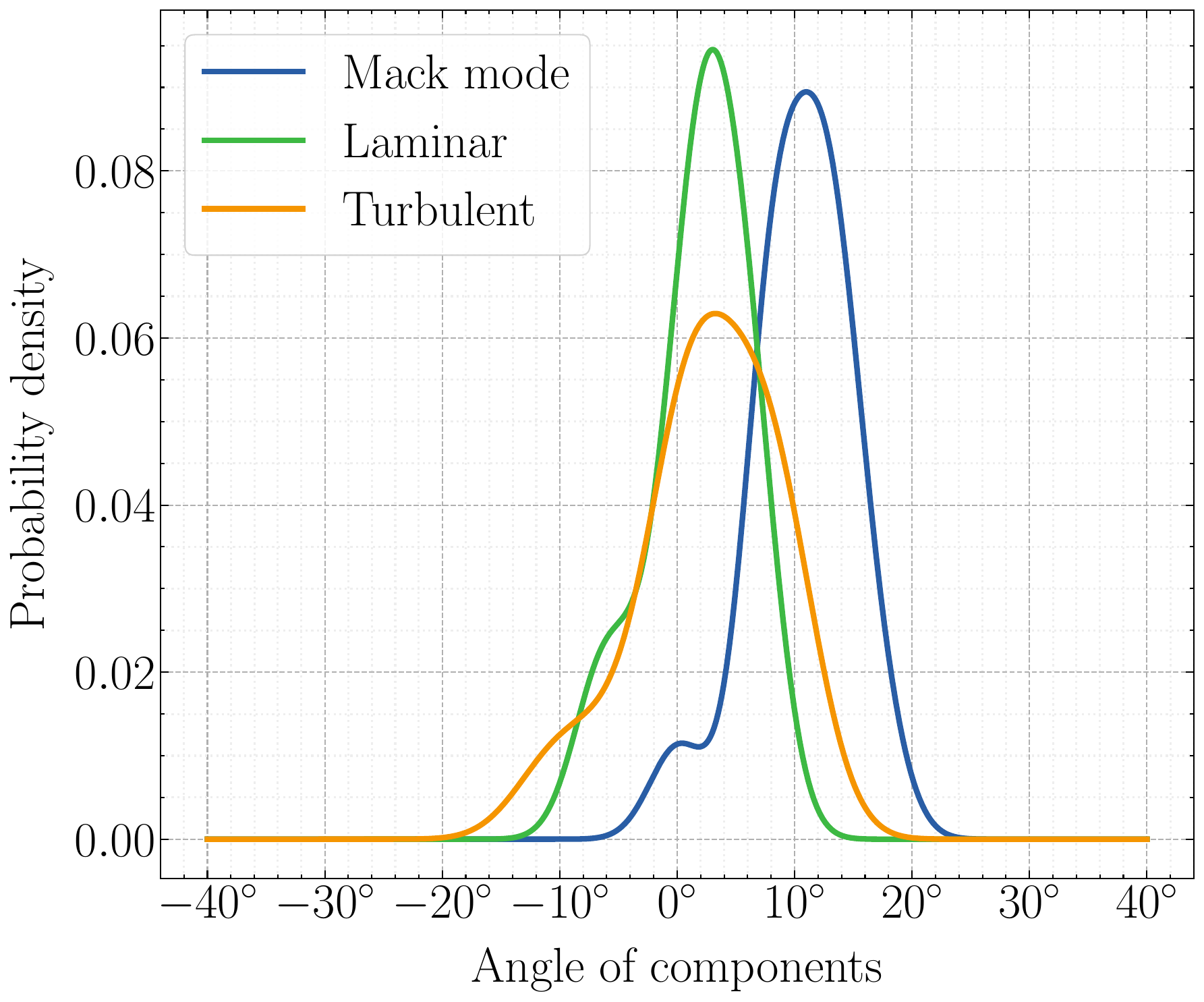}
    \caption{PDFs of the orientation angles for connected components shown in Fig.~\ref{fig:sub3}, Fig.~\ref{fig:Laminarsub2}, and Fig.~\ref{fig:Turbsub4}, with the corresponding flow disturbance state labeled as ``Mack mode'', ``laminar'', and ``turbulent'', respectively.}
    \label{fig:PDF}
\end{figure}

To determine the orientation of each connected component, we perform principal component analysis (PCA) on the node coordinates to identify the direction of maximum variance and calculate its corresponding angle. The process begins by computing the centroid of each component, defined as the mean values of the \(s\) and \(n\) coordinates of its nodes. Then, the coordinates \(s\) and \(n\) of all nodes in the component are worked out with respect to the centroid by simply subtracting the centroid coordinates from the node coordinates. Using these centered coordinates, a data matrix is constructed for each component, where one row contains the mean-subtracted \(s\)-coordinates and the other row contains the mean-subtracted \(n\)-coordinates. From this matrix, we compute the covariance matrix, followed by its eigenvalues and the corresponding eigenvectors. The eigenvector associated with the largest eigenvalue represents the major axis of the component, aligning with the direction of maximum variance. Finally, the orientation angle of this major axis is calculated with respect to the positive \(s\) -axis, providing the average orientation of the component. This method is applied to determine the orientation angles of the connected components shown in Fig.~\ref{fig:sub3}, Fig.~\ref{fig:Laminarsub2}, and Fig.~\ref{fig:Turbsub4}, and the results are shown in the form of a color bar in those respective figures. The probability density functions (PDFs) of the orientation angles from Fig.~\ref{fig:sub3}, Fig.~\ref{fig:Laminarsub2}, and Fig.~\ref{fig:Turbsub4} are derived using using Gaussian Kernel Density Estimation (KDE), and the results are shown in Fig.~\ref{fig:PDF}. The figure shows that the most probable orientation angle of connected components for the schlieren image with rope-like structures is higher than those for laminar flow and flow breakdown images. This trend is also visually evident in Figs.~\ref{fig:sub3}, \ref{fig:Laminarsub2}, and \ref{fig:Turbsub4}.

\begin{figure*}[h!]
    \centering
    \begin{subfigure}[b]{0.8\textwidth}
        \centering
        \includegraphics[width=\textwidth, trim=0cm 12.57cm 0cm 0cm, clip]{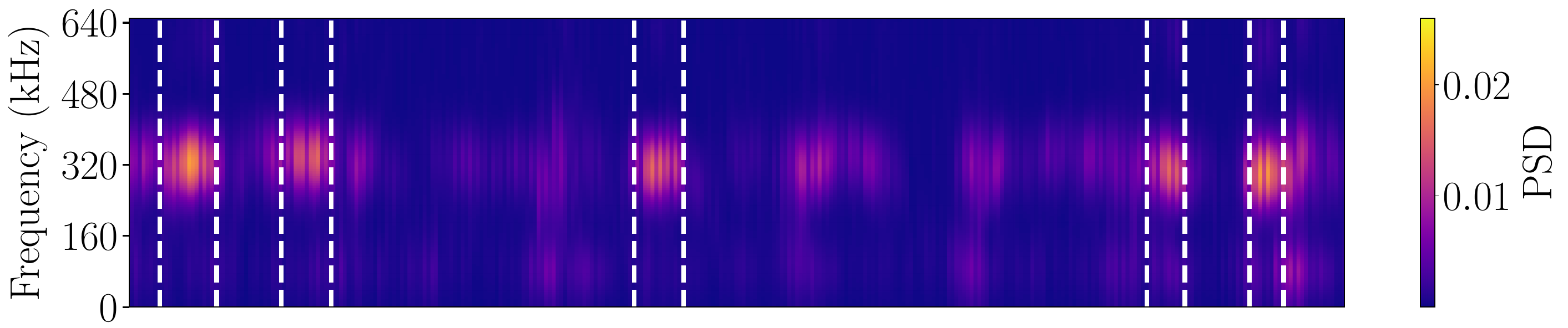}
        \caption{}
        \label{fig:local_frequency_spectra}
    \end{subfigure}
    
    \begin{subfigure}[b]{0.8\textwidth}
        \centering
        \includegraphics[width=\textwidth, trim=0cm 0cm 0cm 9.87cm, clip]{Spectra_and_comp_mode_angles.pdf}
        \caption{}
        \label{fig:most_probable_angles}
    \end{subfigure}
    
    \caption{(a) Temporally-local frequency spectra and (b) time series of the most probable angles of the top connected components over a representative temporal window of 1.6 ms.}
    \label{fig:most_probable_angles_with_time_varying_freq_spectra}
\end{figure*}

To further substantiate the above observations, we obtain a time series of the most probable orientation angle of the connected components derived from all images in the schlieren dataset, and compare it with the temporally-local frequency spectra (obtained using the method outlined in section~\ref{sec:TLFS}). To generate the time series data, a network is constructed for each schlieren image in the dataset. For each network, the top 15 components, based on size, are identified, their orientation angles are assessed, a probability density function (PDF) for these angles is established, and the angle with the highest probability is determined. This approach yields a time series of the most probable angle, with each value in the series corresponding to a particular image and its associated capture time. Representative results from this analysis are presented in Fig.~\ref{fig:most_probable_angles}, wherein the time series of the most probable orientation angle is shown as a color strip over a temporal window of 1.6 ms. The temporally-local frequency spectra for the same window is shown in Fig.~\ref{fig:local_frequency_spectra}. At time instants where the temporally-local frequency spectra indicates the presence of Mack-mode instability, i.e., when the spectral power is high within the frequency range of \SI{300}{\kilo\hertz} to \SI{350}{\kilo\hertz}, the color strip in Fig.~\ref{fig:most_probable_angles} consistently shows that the orientation angle is higher as compared to other time instants where Mack-mode instability is absent in the flow field. It is noted that the time series of the most probable orientation angle also reflects the intermittent nature of the Mack-mode instability.

\begin{figure}[hbt!]
    \centering
        \includegraphics[width=2.6in]{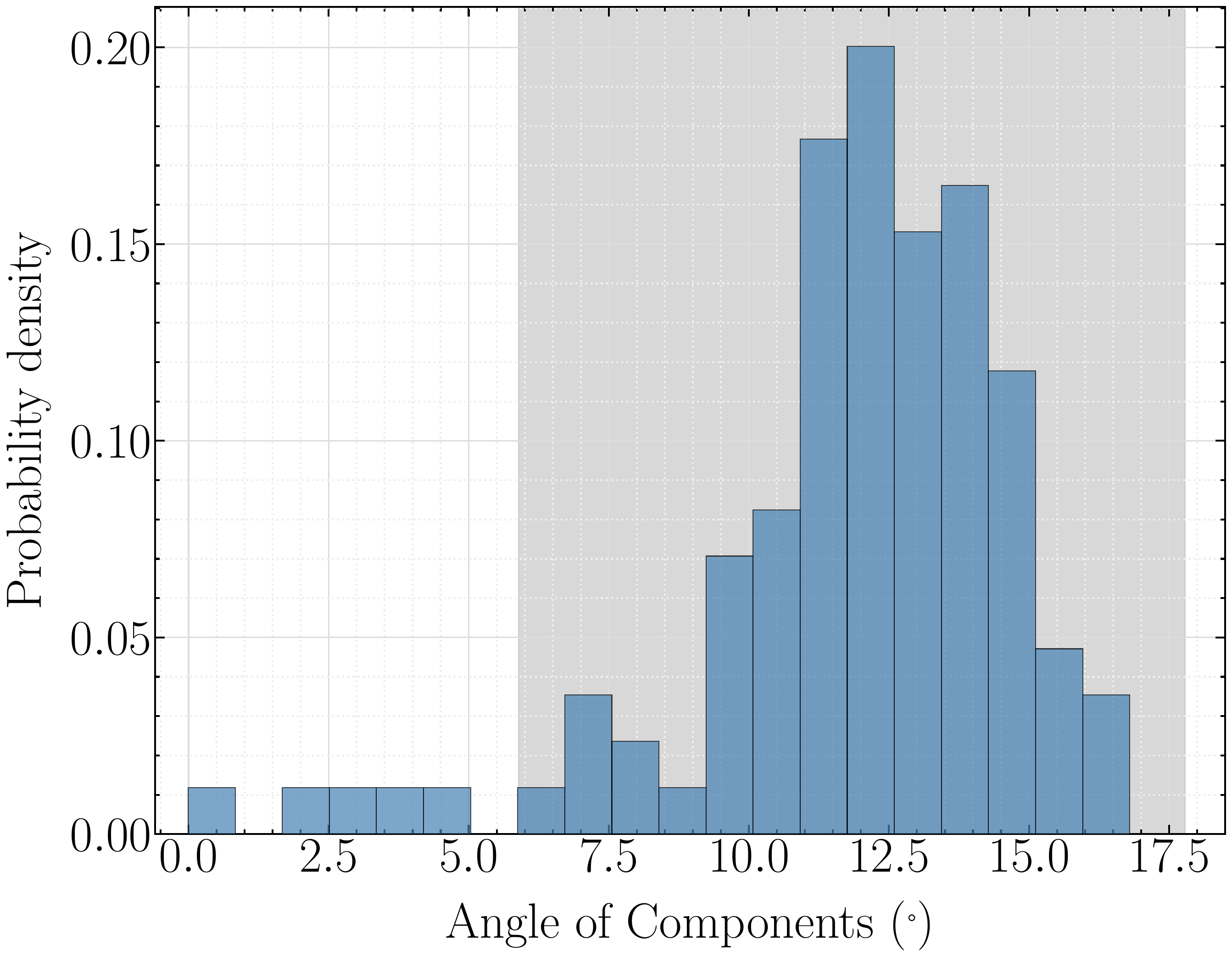}
    \caption{A histogram of most probable component orientation angle at time instants where strong Mack mode activity was recorded. The region shaded in gray indicates the 95\% confidence interval.}
    \label{fig:Mackmode_angles}
\end{figure}

To rigorously quantify the range of nominal Mack mode structure angle, we analyze the most probable angles at time instants where the power spectral density (PSD) indicates strong activity in the frequency band \SI{300}{\kilo\hertz} to \SI{350}{\kilo\hertz}. The criterion for identifying such time instants is set in the following manner. For every time instant in the dataset a representative PSD value for the \SI{300}{\kilo\hertz} to \SI{350}{\kilo\hertz} is obtained by simply averaging the PSD in that band. Choosing an average PSD threshold of 0.01, which is approximately ten times higher than the average PSD value for undisturbed flow, time instants at which the band averaged PSD value exceeds the threshold are identified as occurrences of strong Mack mode activity. A PDF of the most probable orientation angle of the connected components at these time instants is shown in Fig.~\ref{fig:Mackmode_angles}. The 95\% confidence interval for the orientation angle of the connected component is defined by \( \mu \pm 2\sigma \), where \( \mu \) is the mean and \( \sigma \) is the standard deviation of the angle. For our data, \( \mu \pm 2\sigma \) corresponds to an angle range between \SI{5.9}{\degree} and \SI{17.8}{\degree}. Thus, the Mack mode wavepackets in the present study are found to exhibit an average structure angle in the range of \SI{5.9}{\degree} and \SI{17.8}{\degree}, which can be used as a criterion to identify the instants of second mode activity.

\subsubsection{Wavelength of the Mack-mode instability waves}\label{sec:network_wavelength}
\begin{figure*}[hbt!]
    \centering
    \includegraphics[width=\textwidth]{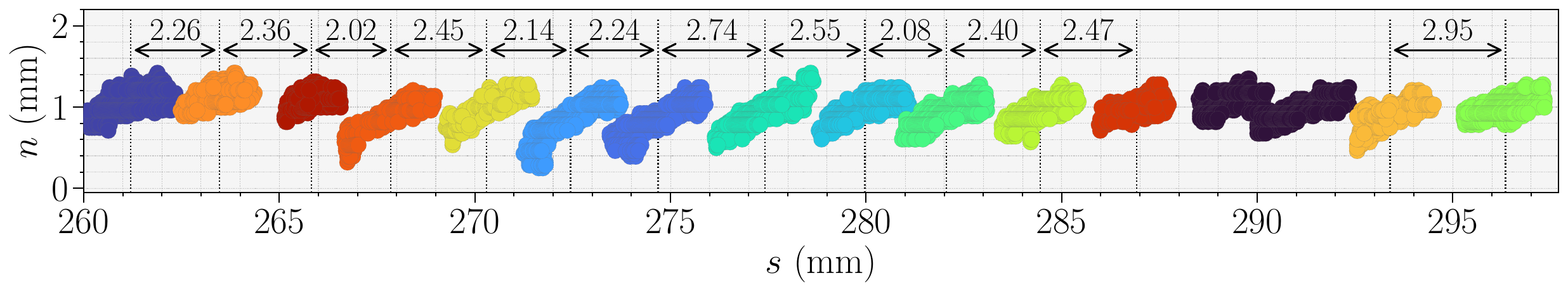}
    \caption{Streamwise distances between centroids of consecutive components with orientation angles between \SI{5.9}{\degree} and \SI{17.8}{\degree} for the schlieren image show in Fig.~\ref{fig:sub1}.}
    \label{fig:Wave_len_comp}
\end{figure*}

\begin{figure}[hbt!]
    \centering
    \includegraphics[width=2.6in]{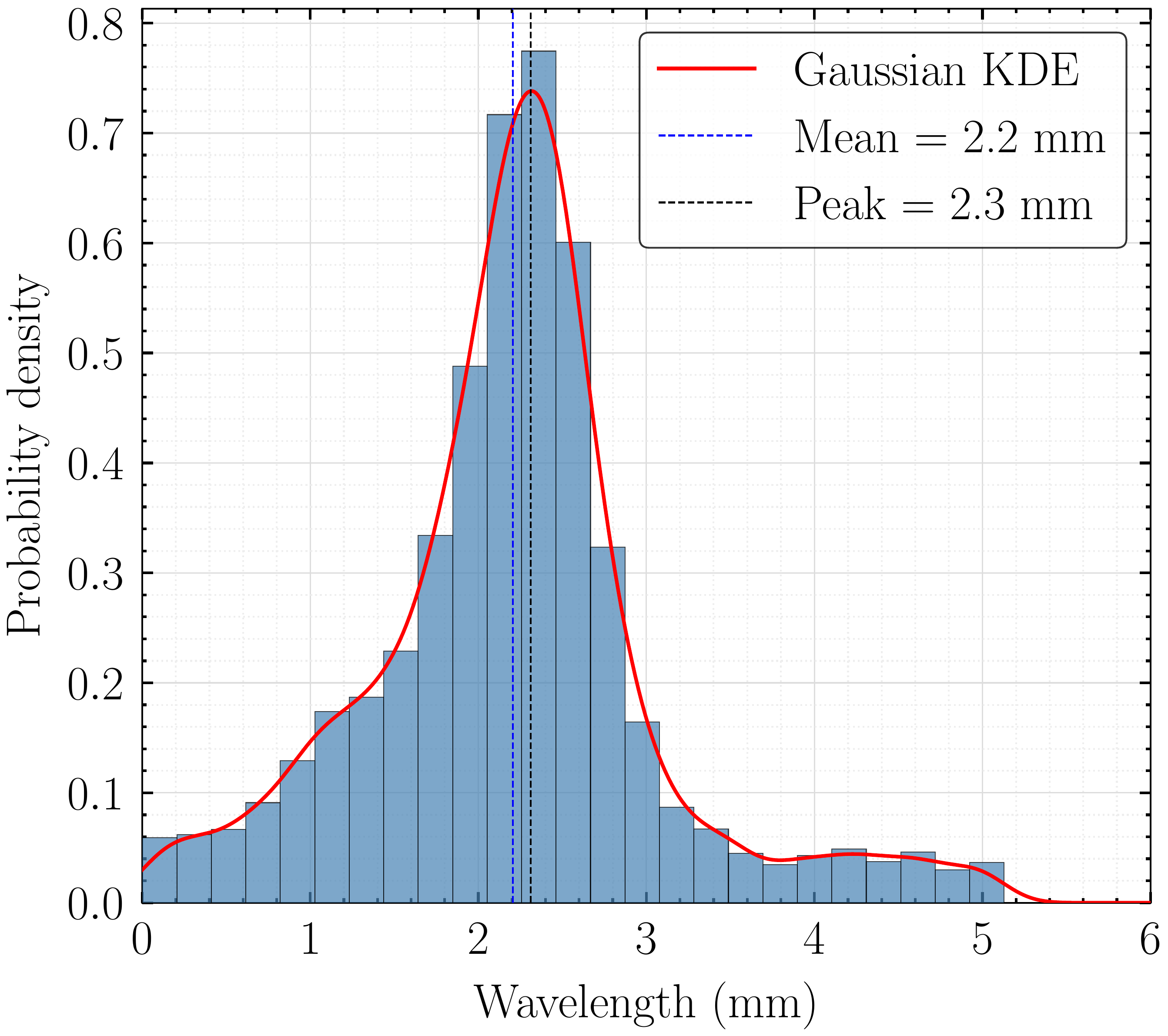}
    \caption{A histogram and Gaussian KDE of the Mack mode wavelengths estimated from all schlieren data.}
    \label{fig:Wave_len_hist}
\end{figure}

We now obtain a statistical estimate for the Mack-mode instability wavelength from the spatial-proximity networks and component analysis. The 15 largest components in each schlieren image are considered and the streamwise locations of their centroids ($s_{\text{mean}}$) are calculated. For each pair of consecutive components with orientation angles between \SI{5.9}{\degree} and \SI{17.8}{\degree}, the streamwise distance between the centroids provides a measure of the wavelength. As a representative example, Fig.~\ref{fig:Wave_len_comp} shows the distances between centroids of consecutive components for the schlieren image shown in Fig.~\ref{fig:sub1}.
We repeat the process for all the schlieren images to obtain a statistical estimate of the wavelength. Figure \ref{fig:Wave_len_hist} shows a histogram of all estimated wavelengths, along with the probability density function (PDF) derived using Gaussian KDE. Outliers were removed from the data by considering values only within the $95\%$ confidence interval. The PDF peak, observed at $2.3 \, \text{mm}$, represents the estimated wavelength of the Mack-mode instability waves, with $0.85 \, \text{mm}$ being the standard deviation. This wavelength corresponds to a wavenumber of 434.8 \(\text{m}^{-1}\), which is in good agreement with the wavenumber estimate of 430 \(\text{m}^{-1}\) obtained earlier using Fourier analysis (see section~\ref{sec:wavenumber_spectra}).

\subsubsection{Velocity of the Mack-mode instability waves}
We now obtain a statistical estimate for the propagation velocity of wavepackets associated with the Mack-mode instability by analyzing components in successive schlieren images. Components with orientation angles between \SI{5.9}{\degree} and \SI{17.8}{\degree} are identified in a particular schlieren image, and their downstream convection is tracked by analysis of the image in conjunction with the schlieren image obtained at the next time instant. Two physical criteria are used to correctly correlate components between the successive images. Firstly, the centroid of any component can only move downstream from the first to the second image. Secondly, a component cannot convect downstream at velocities larger than the freestream flow velocity. The second criterion implies that the distance between the centroids should be less than the maximum possible displacement ($\Delta s_{\text{max}}$), which is obtained from the flow condition at the boundary layer edge as
\[ \Delta s_{\text{max}} = M_e \times \sqrt{\gamma R T_e} \times \Delta t\,, \]
where \( M_e \) and \( T_e \) are Mach number and temperature, respectively, at the boundary layer edge, \( \gamma \) is the ratio of specific heats, \( R \) is the specific gas constant, and \( \Delta t \) is the time interval between two successive images. \( M_e \) and \( T_e \) are obtained using the inviscid Taylor-Maccoll solution \cite{anderson1990modern}. These two criteria mitigate the possibility of misidentification of components due to the intermittent nature of Mack mode disturbances. Further, unambiguous identification of any particular component across two successive images can be achieved when the maximum possible displacement of the component $\Delta s_{\text{max}}$ is smaller than the wavelength of the Mack mode disturbances, estimated to be 2.3 mm in section~\ref{sec:network_wavelength}. For the present schlieren dataset, image pairs with \SI{2.3}{\micro\second} time interval satisfy this condition; \(\Delta s_{\text{max}} = 1.92\, \text{mm}\) when \(\Delta t\) = \SI{2.3}{\micro\second} in the above equation. Hence, only the image pairs with \SI{2.3}{\micro\second} time interval are used from the pulse-burst schlieren data for estimating the propagation velocity.

\begin{figure*}[hbt!]
    \centering
    \includegraphics[width=\textwidth]{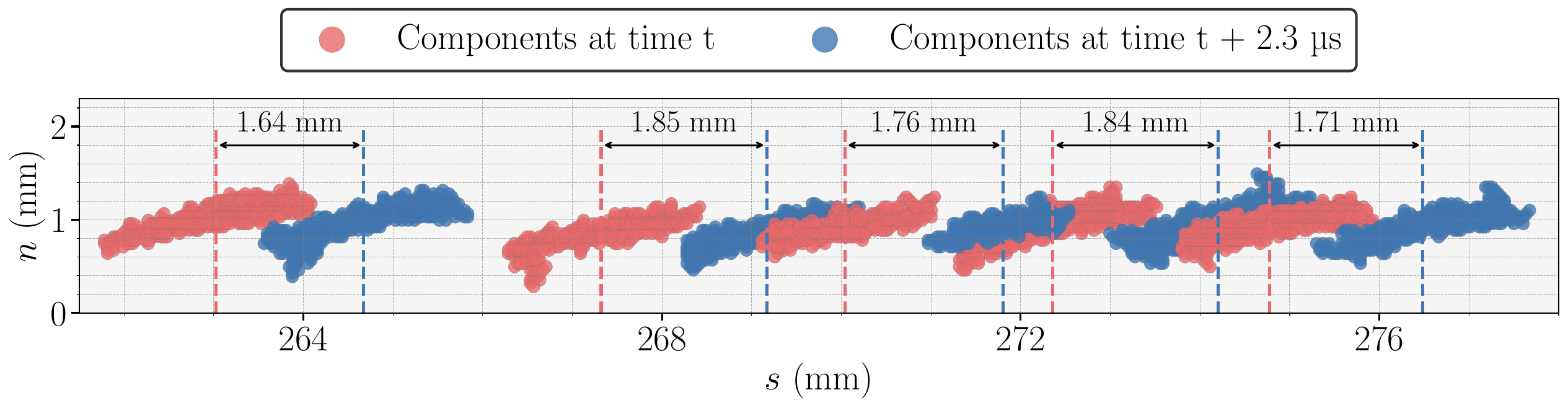}
    \caption{Streamwise displacements \(\Delta s\) of components between a pair of schlieren images with a \SI{2.3}{\micro\second} interval.}
    \label{fig:velocity_disp_schem}
\end{figure*}

\begin{figure}[hbt!]
    \centering
    \includegraphics[width=2.6in]{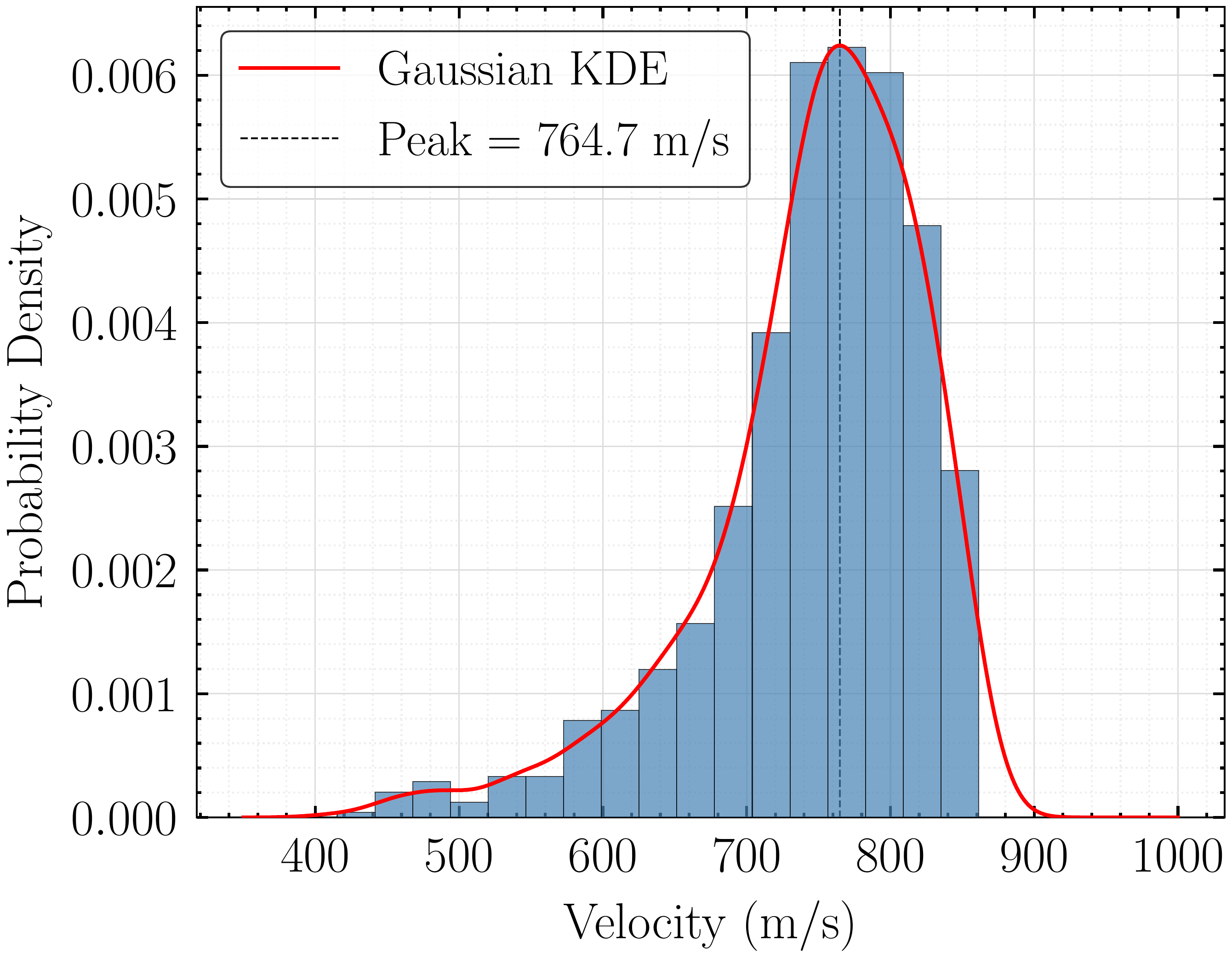}
    \caption{A histogram of propagation velocities of the Mack mode components.}
    \label{fig:velocity_hist}
\end{figure}

As a representative example, Fig.~\ref{fig:velocity_disp_schem} shows five different convecting components identified in a particular pair of schlieren images with a \SI{2.3}{\micro\second} time interval. The displacements of the component centroids are marked in the figure. The velocity of each component is then simply written as a ratio of its displacement and the time interval between the successive images (\emph{i.e.}, \SI{2.3}{\micro\second}). Velocity of components are obtained from this procedure for all the \SI{2.3}{\micro\second} time interval image pairs in the schlieren data, and the resulting velocity distribution is shown in Fig.~\ref{fig:velocity_hist}. The velocity PDF peaks at \( 764 \, \text{m/s} \), which matches closely with the estimated PDF peak of 763.9 m/s obtained using image cross-correlation techniques (see section~\ref{sec:wave_propogation_speeds}). The mean propagation velocity estimated using the network analysis method is \( 746.7 \pm 76.21 \, \text{m/s} \) (95\% confidence interval), while the image cross-correlation techniques yielded a mean propagation velocity of \( 762.7 \pm 12.2 \, \text{m/s} \) (95\% confidence interval). The difference of 17.3 m/s between the mean and the modal peak velocities obtained using the network analysis-based method is attributed to the asymmetry of the velocity distribution, which is skewed towards lower velocities, resulting in a broader distribution and higher variability in the estimated mean phase speed as compared to the image cross-correlation method. Consequently, the peak of the PDF, rather than the mean, is treated to be a more accurate indicator of the true phase velocity of the Mack-mode instability waves.

\section{Brief Conclusions} \label{sec:conclusions}
A common pathway for hypersonic boundary layer flow transition from a laminar to a turbulent state is via the Mach mode instability, an inviscid instability mechanism of acoustic nature. In this work we characterized the Mack-mode instability by performing experiments on a \( 7^\circ \) cone at zero angle of attack in a Mach 6 flow. High-speed schlieren images of the boundary layer were captured in the flow transition region using the pulse-burst method, which allowed for determining the wavelength and propagation speed of the coherent disturbances generated by the Mack-mode instability using standard Fourier analysis methods.

The schlieren data also enabled the application of modern data-driven methods. With the same dataset, we demonstrated \edit{for the first time} the viability of applying network science in objectively determining the characteristics of Mack-mode instability. Time-varying spatial proximity networks were constructed based on the schlieren light intensity field. The connected components of the network were estimated and utilized in investigating the spatio-temporal dynamics of the flow field. In the flow transition regime, the connected components represent the disturbance wave packets associated with the Mack-mode instability. We find that the angle of the components is a good geometric metric for wave packet identification since the wave packets have a characteristic rope-like structure. The orientation angles of the components, which represent the physical inclination of the rope-like structures, were estimated by performing the principal component analysis of the coordinates of the nodes. From this analysis the inclination angles (with respect to the wall) of the rope-like structures were found to be in the range of $5.9^\circ-17.8^\circ$. We also demonstrated that the spatial-proximity network analysis is an effective framework for estimating the wavelength and propagation velocity of wave packets associated with the Mack-mode instability. The results from network analysis are in good agreement with those obtained from conventional Fourier analysis tools.

While this paper establishes network analysis as an alternate approach for instability characterization, we note that network analysis offers advantages for analysis of flow disturbances that are intermittent in nature. For such cases the utility and applicability of Fourier based methods are limited. \edit{In the context of high-speed boundary flow transition, network analysis can be employed as a powerful tool to obtain new insights into the transition process. And finally,} from an experimental viewpoint, network analysis offers a simplification in experimental methodology since it does not necessarily require schlieren data obtained using the pulse-burst technique, with uniformly sampled data sufficing for analysis.

\section*{Acknowledgments}
This work was supported by a research grant (S.D.) from the DRDO Raman Centre of Excellence at IISc. Partial support was also received from GoI Ministry of Education in the form a PhD Scholarship (S.P.M.) and from GoI Department of Science and Technology in the form of an INSPIRE Faculty Fellowship (N.K.). B.M. Shiva Shankar, N. Shanta Kumar, and M. Harish assisted in the operations and maintenance of the Roddam Narasimha Hypersonic Wind Tunnel facility.

\bibliography{sample}

\begin{thebibliography}{34}
\newcommand{\enquote}[1]{``#1''}
\providecommand{\natexlab}[1]{#1}
\providecommand{\url}[1]{\texttt{#1}}
\providecommand{\urlprefix}{URL }
\expandafter\ifx\csname urlstyle\endcsname\relax
  \providecommand{\doi}[1]{\discretionary{}{}{}https://doi.org/#1}\else
  \providecommand{\doi}[1]{\discretionary{}{}{}\urlstyle{rm}\url{https://doi.org/#1}}\fi

\bibitem[{Fedorov(2011)}]{fedorov2011transition}
Fedorov, A., \enquote{Transition and Stability of High-Speed Boundary Layers,} \emph{Annual Review of Fluid Mechanics}, Vol.~43, No.~1, 2011, pp. 79--95.
\newblock \doi{10.1146/annurev-fluid-122109-160750}.

\bibitem[{Zhong and Wang(2012)}]{zhong2012direct}
Zhong, X., and Wang, X., \enquote{Direct Numerical Simulation on the Receptivity, Instability, and Transition of Hypersonic Boundary Layers,} \emph{Annual Review of Fluid Mechanics}, Vol.~44, No.~1, 2012, pp. 527--561.
\newblock \doi{10.1146/annurev-fluid-120710-101208}.

\bibitem[{Mack(1975)}]{mack1975linear}
Mack, L.~M., \enquote{Linear Stability Theory and the Problem of Supersonic Boundary-Layer Transition,} \emph{AIAA Journal}, Vol.~13, No.~3, 1975, pp. 278--289.
\newblock \doi{10.2514/3.49693}.

\bibitem[{Mack(1987)}]{mack1987review}
Mack, L.~M., \enquote{Review of Linear Compressible Stability Theory,} \emph{Stability of Time Dependent and Spatially Varying Flows: Proceedings of the Symposium on the Stability of Time Dependent and Spatially Varying Flows}, Springer, 1987, pp. 164--187.

\bibitem[{Betchov(2012)}]{betchov2012stability}
Betchov, R., \emph{Stability of Parallel Flows}, Elsevier, 2012.

\bibitem[{Criminale et~al.(2018)Criminale, Jackson, and Joslin}]{criminale2018theory}
Criminale, W.~O., Jackson, T.~L., and Joslin, R.~D., \emph{Theory and Computation in Hydrodynamic Stability}, Cambridge University Press, 2018.

\bibitem[{Mack(1984)}]{mack1984boundary}
Mack, L.~M., \enquote{Boundary-Layer Linear Stability Theory,} \emph{AGARD Rep}, Vol. 709, 1984.

\bibitem[{Reed et~al.(1996)Reed, Saric, and Arnal}]{reed1996linear}
Reed, H.~L., Saric, W.~S., and Arnal, D., \enquote{Linear Stability Theory Applied to Boundary Layers,} \emph{Annual Review of Fluid Mechanics}, , No.~28, 1996, pp. 389--428.
\newblock \doi{10.1146/annurev.fl.28.010196.002133}.

\bibitem[{Laurence et~al.(2012)Laurence, Wagner, Hannemann, Wartemann, L{\"u}deke, Tanno, and Itoh}]{laurence2012time}
Laurence, S.~J., Wagner, A., Hannemann, K., Wartemann, V., L{\"u}deke, H., Tanno, H., and Itoh, K., \enquote{Time-Resolved Visualization of Instability Waves in a Hypersonic Boundary Layer,} \emph{AIAA Journal}, Vol.~50, No.~1, 2012, pp. 243--246.
\newblock \doi{10.2514/1.J051112}.

\bibitem[{Zhang et~al.(2013)Zhang, Tang, and Lee}]{zhang2013hypersonic}
Zhang, C.-H., Tang, Q., and Lee, C.-B., \enquote{Hypersonic Boundary-Layer Transition on a Flared Cone,} \emph{Acta Mechanica Sinica}, Vol.~29, 2013, pp. 48--54.
\newblock \doi{10.1007/s10409-013-0009-2}.

\bibitem[{Laurence et~al.(2014)Laurence, Wagner, and Hannemann}]{laurence2014schlieren}
Laurence, S.~J., Wagner, A., and Hannemann, K., \enquote{Schlieren-Based Techniques for Investigating Instability Development and Transition in a Hypersonic Boundary Layer,} \emph{Experiments in Fluids}, Vol.~55, 2014, pp. 1--17.
\newblock \doi{10.1007/s00348-014-1782-9}.

\bibitem[{Grossir et~al.(2014)Grossir, Pinna, Bonucci, Regert, Rambaud, and Chazot}]{grossir2014hypersonic}
Grossir, G., Pinna, F., Bonucci, G., Regert, T., Rambaud, P., and Chazot, O., \enquote{Hypersonic Boundary Layer Transition on a 7 Degree Half-Angle Cone at Mach 10,} \emph{7th AIAA Theoretical Fluid Mechanics Conference}, 2014, p. 2779.
\newblock \doi{10.2514/6.2014-2779}.

\bibitem[{Kennedy et~al.(2018)Kennedy, Laurence, Smith, and Marineau}]{kennedy2018visualization}
Kennedy, R.~E., Laurence, S.~J., Smith, M.~S., and Marineau, E.~C., \enquote{Visualization of the Second-Mode Instability on a Sharp Cone at Mach 14,} \emph{2018 AIAA Aerospace Sciences Meeting}, 2018, p. 2083.
\newblock \doi{10.2514/6.2018-2083}.

\bibitem[{Siddiqui et~al.(2021)Siddiqui, Gragston, Saric, and Bowersox}]{siddiqui2021mack}
Siddiqui, F., Gragston, M., Saric, W.~S., and Bowersox, R. D.~W., \enquote{Mack-Mode Instabilities on a Cooled Flared Cone with Discrete Roughness Elements at Mach 6,} \emph{Experiments in Fluids}, Vol.~62, 2021, pp. 1--13.
\newblock \doi{10.1007/s00348-021-03304-6}.

\bibitem[{Scholten et~al.(2022)Scholten, Paredes, Hill, Borg, Jewell, and Choudhari}]{scholten2022linear}
Scholten, A., Paredes, P., Hill, J.~L., Borg, M., Jewell, J.~S., and Choudhari, M.~M., \enquote{Linear Instabilities over Ogive-Cylinder Models at Mach 6,} \emph{AIAA Journal}, Vol.~60, No.~8, 2022, pp. 4478--4491.
\newblock \doi{10.2514/1.J061611}.

\bibitem[{Barab{\'a}si(2013)}]{barabasi2013network}
Barab{\'a}si, A.-L., \enquote{Network Science,} \emph{Philosophical Transactions of the Royal Society A: Mathematical, Physical and Engineering Sciences}, Vol. 371, No. 1987, 2013, p. 20120375.
\newblock \doi{10.1098/rsta.2012.0375}.

\bibitem[{Newman(2018)}]{newman2018networks}
Newman, M., \emph{Networks}, Oxford University Press, 2018.
\newblock \doi{10.1093/oso/9780198805090.001.0001}.

\bibitem[{Van~Steen(2010)}]{van2010graph}
Van~Steen, M., \emph{Graph Theory and Complex Networks: An Introduction}, Self-published, Maastricht, The Netherlands, 2010.

\bibitem[{Nair and Taira(2015)}]{nair2015network}
Nair, A.~G., and Taira, K., \enquote{Network-Theoretic Approach to Sparsified Discrete Vortex Dynamics,} \emph{Journal of Fluid Mechanics}, Vol. 768, 2015, pp. 549--571.
\newblock \doi{10.1017/jfm.2015.97}.

\bibitem[{Taira et~al.(2016)Taira, Nair, and Brunton}]{taira2016network}
Taira, K., Nair, A.~G., and Brunton, S.~L., \enquote{Network Structure of Two-Dimensional Decaying Isotropic Turbulence,} \emph{Journal of Fluid Mechanics}, Vol. 795, 2016, p.~R2.
\newblock \doi{10.1017/jfm.2016.235}.

\bibitem[{Meena and Taira(2021)}]{meena2021identifying}
Meena, M.~G., and Taira, K., \enquote{Identifying Vortical Network Connectors for Turbulent Flow Modification,} \emph{Journal of Fluid Mechanics}, Vol. 915, 2021, p. A10.
\newblock \doi{10.1017/jfm.2021.35}.

\bibitem[{Taira and Nair(2022)}]{taira2022network}
Taira, K., and Nair, A.~G., \enquote{Network-Based Analysis of Fluid Flows: Progress and Outlook,} \emph{Progress in Aerospace Sciences}, Vol. 131, 2022, p. 100823.
\newblock \doi{10.1016/j.paerosci.2022.100823}.

\bibitem[{Iacobello et~al.(2021)Iacobello, Ridolfi, and Scarsoglio}]{iacobello2021review}
Iacobello, G., Ridolfi, L., and Scarsoglio, S., \enquote{A Review on Turbulent and Vortical Flow Analyses via Complex Networks,} \emph{Physica A: Statistical Mechanics and Its Applications}, Vol. 563, 2021, p. 125476.
\newblock \doi{10.1016/j.physa.2020.125476}.

\bibitem[{Iacobello et~al.(2019)Iacobello, Scarsoglio, Kuerten, and Ridolfi}]{iacobello2019lagrangian}
Iacobello, G., Scarsoglio, S., Kuerten, J. G.~M., and Ridolfi, L., \enquote{Lagrangian Network Analysis of Turbulent Mixing,} \emph{Journal of Fluid Mechanics}, Vol. 865, 2019, pp. 546--562.
\newblock \doi{10.1017/jfm.2019.79}.

\bibitem[{Chopra et~al.(2024)Chopra, Mittal, and Sujith}]{chopra2024evolution}
Chopra, G., Mittal, S., and Sujith, R.~I., \enquote{Evolution of Clusters of Turbulent Reattachment Due to Shear Layer Instability in Flow Past a Circular Cylinder,} \emph{Physics of Fluids}, Vol.~36, No.~1, 2024.
\newblock \doi{10.1063/5.0187414}.

\bibitem[{Krishnan et~al.(2019)Krishnan, Sujith, Marwan, and Kurths}]{krishnan2019emergence}
Krishnan, A., Sujith, R.~I., Marwan, N., and Kurths, J., \enquote{On the Emergence of Large Clusters of Acoustic Power Sources at the Onset of Thermoacoustic Instability in a Turbulent Combustor,} \emph{Journal of Fluid Mechanics}, Vol. 874, 2019, pp. 455--482.
\newblock \doi{10.1017/jfm.2019.429}.

\bibitem[{Thasu and Duvvuri(2024)}]{thasu2024measurement}
Thasu, P.~S., and Duvvuri, S., \enquote{Measurement of Freestream Noise in a Hypersonic Wind Tunnel,} \emph{Experiments in Fluids}, Vol.~65, No.~4, 2024, p.~45.
\newblock \doi{10.1007/s00348-024-03783-3}.

\bibitem[{Premika(2024)}]{premika2024self}
Premika, T.~S., \enquote{Self-Sustained Oscillations in High-Speed Cylinder Wake Flows,} Ph.{D}. {D}issertation, Indian Institute of Science, Bengaluru, India, 2024.

\bibitem[{Sasidharan and Duvvuri(2021)}]{sasidharan2021large}
Sasidharan, V., and Duvvuri, S., \enquote{Large- and Small-Amplitude Shock-Wave Oscillations over Axisymmetric Bodies in High-Speed Flow,} \emph{Journal of Fluid Mechanics}, Vol. 913, 2021, p.~R7.
\newblock \doi{10.1017/jfm.2021.115}.

\bibitem[{Thasu and Duvvuri(2022)}]{thasu2022strouhal}
Thasu, P.~S., and Duvvuri, S., \enquote{Strouhal Number Universality in High-Speed Cylinder Wake Flows,} \emph{Physical Review Fluids}, Vol.~7, No.~8, 2022, p. L081401.
\newblock \doi{10.1103/PhysRevFluids.7.L081401}.

\bibitem[{Kumar et~al.(2024)Kumar, Sasidharan, Kumara, and Duvvuri}]{kumar2024model}
Kumar, G., Sasidharan, V., Kumara, A.~G., and Duvvuri, S., \enquote{A Model for Frequency Scaling of Flow Oscillations in High-Speed Double Cones,} \emph{Journal of Fluid Mechanics}, Vol. 988, 2024, p. A37.
\newblock \doi{10.1017/jfm.2024.449}.

\bibitem[{Laurence et~al.(2016)Laurence, Wagner, and Hannemann}]{laurence2016experimental}
Laurence, S.~J., Wagner, A., and Hannemann, K., \enquote{Experimental Study of Second-Mode Instability Growth and Breakdown in a Hypersonic Boundary Layer Using High-Speed Schlieren Visualization,} \emph{Journal of Fluid Mechanics}, Vol. 797, 2016, pp. 471--503.
\newblock \doi{10.1017/jfm.2016.280}.

\bibitem[{Parziale et~al.(2015)Parziale, Shepherd, and Hornung}]{parziale2015observations}
Parziale, N.~J., Shepherd, J.~E., and Hornung, H.~G., \enquote{Observations of Hypervelocity Boundary-Layer Instability,} \emph{Journal of Fluid Mechanics}, Vol. 781, 2015, pp. 87--112.
\newblock \doi{10.1017/jfm.2015.489}.

\bibitem[{Anderson(1990)}]{anderson1990modern}
Anderson, J.~D., \emph{Modern Compressible Flow: With Historical Perspective}, McGraw-Hill, New York, 1990.

\end{thebibliography}
\end{document}